\documentclass[
reprint,
amsmath,amssymb,
aps,
]{revtex4-2}

\usepackage{graphicx}
\usepackage{dcolumn}
\usepackage{bm}
\usepackage{hyperref}
\usepackage{subcaption}
\usepackage{relsize}
\usepackage{cleveref}
\usepackage{physics}
\usepackage{IEEEtrantools}
\usepackage{qcircuit}
\usepackage{bbm}
\usepackage{float} 

\newcommand{\outputgroupv}[6]%
{\POS"#1,#3"."#2,#3"."#1,#3"."#2,#3"!C*+<#4>\frm{\}}, \POS"#1,#3"."#2,#3"."#1,#3"."#2,#3"*!C!<-2em,#5>=<0em>{#6}}

\newcommand{\transp}{{\mathsmaller T}}

\newcommand{\eqbox}[3]{\begin{IEEEeqnarraybox}[][#1]{#2}#3\end{IEEEeqnarraybox}}

\newcommand{\hcos}[1]{\cos\qty\Big(\frac{#1}{2})}
\newcommand{\hsin}[1]{\sin\qty\Big(\frac{#1}{2})}

\begin{document}
	
	\title{Bell Diagonal and Werner state generation: 
		entanglement, \\ 
		non-locality, steering and discord on the IBM quantum computer}
	
	\author{Elias Riedel G\aa rding}
	\altaffiliation[Also at ]{Department of Physics, Royal Institute of Technology (KTH), Stockholm, Sweden.}
	\affiliation{Institute of Physics, \'{E}cole Polytechnique F\'{e}d\'{e}rale de Lausanne (EPFL), Lausanne, CH-1015, Switzerland}
	\author{Nicolas Schwaller}
	\affiliation{Institute of Physics, \'{E}cole Polytechnique F\'{e}d\'{e}rale de Lausanne (EPFL), Lausanne, CH-1015, Switzerland}
	\author{Chun Lam Chan}
	\affiliation{Laboratoire de Th\'{e}orie des Communications, Facult\'{e} Informatique et Communications, \'{E}cole Polytechnique F\'{e}d\'{e}rale de Lausanne (EPFL), Lausanne, CH-1015, Switzerland}
	\author{Su Yeon Chang}
	\affiliation{Institute of Physics, \'{E}cole Polytechnique F\'{e}d\'{e}rale de Lausanne (EPFL), Lausanne, CH-1015, Switzerland}
	\author{Samuel Bosch}
	\altaffiliation[Also at ]{Integrated Systems Laboratory, \'{E}cole Polytechnique F\'{e}d\'{e}rale de Lausanne (EPFL), Lausanne, CH-1015, Switzerland.}
	\affiliation{Institute of Physics, \'{E}cole Polytechnique F\'{e}d\'{e}rale de Lausanne (EPFL), Lausanne, CH-1015, Switzerland}
	\author{Frederic Gessler}
	\affiliation{Laboratoire de Th\'{e}orie des Communications, Facult\'{e} Informatique et Communications, \'{E}cole Polytechnique F\'{e}d\'{e}rale de Lausanne (EPFL), Lausanne, CH-1015, Switzerland}
	\author{Willy Robert Laborde}
	\affiliation{Institute of Physics, \'{E}cole Polytechnique F\'{e}d\'{e}rale de Lausanne (EPFL), Lausanne, CH-1015, Switzerland}
	\author{Javier Naya Hernandez}
 	\altaffiliation[Also at ]{School of Science and Engineering, Tecnol\'{o}gico de Monterrey, Monterrey, Mexico}
	\author{Xinyu Si}
	\affiliation{Institute of Physics, \'{E}cole Polytechnique F\'{e}d\'{e}rale de Lausanne (EPFL), Lausanne, CH-1015, Switzerland}
	\author{Marc-Andr\'{e} Dupertuis}
	\email{marc-andre.dupertuis@epfl.ch}
	\affiliation{Institute of Physics, \'{E}cole Polytechnique F\'{e}d\'{e}rale de Lausanne (EPFL), Lausanne, CH-1015, Switzerland}
	\author{Nicolas Macris}
	\email{nicolas.macris@epfl.ch}
	\affiliation{Laboratoire de Th\'{e}orie des Communications, Facult\'{e} Informatique et Communications, \'{E}cole Polytechnique F\'{e}d\'{e}rale de Lausanne (EPFL), Lausanne, CH-1015, Switzerland}

	\date{\today}
	
	\begin{abstract}
	We propose the first correct special-purpose quantum circuits for preparation of Bell-diagonal states (BDS), and implement them on the IBM Quantum computer, characterizing and testing complex aspects of their quantum correlations in the full parameter space. 
	Among the circuits proposed, one involves only two quantum bits but requires adapted quantum tomography routines handling classical bits in parallel. The entire class of Bell-diagonal states is generated, and a number of characteristic indicators, namely entanglement of formation and concurrence, CHSH non-locality, steering and discord, are experimentally evaluated over the full parameter space and compared with theory. As a by-product of this work we also find a remarkable general inequality between ``quantum discord'' and ``asymmetric relative entropy of discord'': the former never exceeds the latter. We also prove that for all BDS the two coincide.
	\end{abstract}
	
	\maketitle
	
	\section{Introduction} \label{sec:level1}
	
	The field of quantum computing offers an entirely new paradigm of computation which promises significant asymptotic speedups over classical computers for certain
	problems \cite{Shor_1995, Grover_1996, QCircuit_WaveF_Chemistry, Quantum_Chemistry_review}
	as well as new kinds of highly secure cryptographic protocols \cite{QCrypt_Gisin2002, Short_Review_QCryptography, Quanta57}.
	At the foundation of this new field lies the theory of quantum information which, among other things,
	provides insight into the structure of the state space of a system of many qubits as well as
	ways to characterize and mitigate noise.
	Technological progress is impressive, providing publicly available programmable quantum platforms with a dozen qubits like IBM Quantum Experience (IBM Q, see \cite{IBMQ2019}). Recently a team at Google and NASA demonstrated the thrilling superior performance of a 53 qubit quantum processor called Sycamore \cite{SycamoreX2019}, and claimed to achieve quantum supremacy (see also \cite{Pednault2019,Aaronson2019,Preskill2019} for criticisms).

Bell states are archetypal examples of entangled two-qubit pure quantum states. Statistical mixtures of Bell states are called Bell-diagonal states (BDS). They form a very interesting restricted class of states which, despite their relative simplicity, display a rich variety of correlations, and have played a crucial role in the theory of quantum information. Because they form a representative three-dimensional subspace of the full $15$-dimensional space of two-qubit mixed states, they are often used as a testing ground for measures of quantum correlation, such as as entropic measures \cite{Entropic_measures} or quantum discord \cite{Quantum_Discord_Ollivier_Zurek2001}. Indeed, progress in quantum information theory led physicists to think about measures of quantum correlations beyond entanglement \cite{Bennett-et-al2001}. During the last decades the nature of entanglement has been the subject of an ever increasing number of studies, not only because of its intriguing nature related to Bell inequality violations, but also because of its formerly unsuspected complexity, in particular concerning quantum mixtures \cite{QEntanglementReview}. 
The concept of entanglement which seems trivial for bipartite pure states was found difficult to characterize for quantum mixtures, because of the lack of universal entanglement measure \cite{QEntanglementReview}. Further daunting complexities were found in the case with more than two parties, since inequivalent classes of entanglement under LOCC (Local Operations and Classical Communication) manipulation could be defined \cite{GHZWentanglement}. Further surprises came when the encompassing subject of quantum versus classical correlations was found to be distinct from the entanglement/separability paradigm, and various notions of discord were introduced \cite{Quantum_Discord_Ollivier_Zurek2001,henderson2001classical,Modi2010, Modi2012,ferraro2010almost,Discord_review}. Indeed separable mixed states can still exhibit useful quantum correlations, even for only two parties. Another previously overlooked concept is steering, the property to steer a quantum state from another location, and it was found to be an even more subtle notion (precisely formalized for the first time in \cite{Wiseman-et-al2007,Jones-et-al2007}). Steering is intermediate between non-separability and Bell-non-locality, and has duly attracted considerable attention (see \cite{uola2019quantum} for a recent review). A rewarding consequence of all these discoveries about entanglement and quantum correlations is that most of them prove useful for specific quantum tasks \cite{QEntanglementReview,Discord_review}.

BDS are especially interesting in the context of calculating quantum correlations because these in many cases have an analytical expression. For instance, the ``quantum discord'' of BDS and so-called $X$ states \cite{Luo2008, Analytical_Discord_BDS, Analytical_Discord_BDS_2, Analytical_Discord_BDS_3} has been calculated, and, as it will be shown, for BDS it coincides with ``asymmetric relative entropy of discord''. The computation of such correlations is essential in quantum information theory, to classify systems according to the extent they exhibit non-classical behavior. In particular, one application of these ideas is the problem of witnessing non-classicality of inaccessible objects \cite{Revealing_non_classicality}.
Moreover, experimental computations of such correlations for BDS have been reported. For instance  \cite{QCorr_Experimental} and \cite{QDiscord_Exp} detected a certain amount of quantum discord in magnetic resonance experiments, evidencing the existence of non-classical correlations without entanglement. Another type of non-classical correlations, firstly quantum steering, has been observed experimentally \cite{Exp_Steering_Sun,Exp_Steering_Wittmann_2012},
as well as negativity of entanglement \cite{Measuring_QCor_Silva2013}.

We propose here the first correct special-purpose quantum circuit for preparation of any arbitrary BDS on quantum computers. Indeed the previous proposal of Pozzobom and Maziero \cite{Pozzobom2019} falls clearly short of this goal, since it is impossible to cover the intended three-dimensional space of target states with only two parameters. In this work, we present two new circuits, either of which enables to prepare any BDS, and provide implementations in Qiskit \cite{Qiskit}. Furthermore, while the original circuit \cite{Pozzobom2019} uses four qubits, we show how the task can also be accomplished with only two qubits using unread measurements. The latter requires certain amendments to the standard quantum tomography procedure in Qiskit. The two-qubit circuits rely on post-measurement gates and classically conditioned measurements, which are currently unsupported on IBM Q devices. As such, they highlight the importance of continuing the development of hardware features for quantum computers.

The paper is structured as follows.
After introducing BDS and its relevant subset called Werner states (WS) in \cref{sect:BDS_WS_intro}, we propose a parameterization of the whole set of BDS allowing their generation by four-qubit and two-qubit circuits in \cref{sect:circuits}. \Cref{sec:Qiskitimplementation} details the implementation of the circuits with Qiskit \cite{Qiskit}, notably for two qubit circuits which require new tomography functions.
Entanglement of formation and concurrence, CHSH non-locality, steering and discord are reviewed and reexamined in \cref{sect:entanglement_discord}, and visualized for BDS.  In the last section \cref{sect:IBMQ_results} we study on the IBM Q platform the achievable fidelity for Werner states, as well as classical correlations, mutual information and discord for BDS. Results of simulations on the IBM Q simulator with noise models for real devices, as well experimental results on real devices are reported and discussed.

\section{Bell Diagonal and Werner states and their properties}
\label{sect:BDS_WS_intro}
	
Bell states are defined as maximally entangled basis states of the two-qubit Hilbert space $\mathcal{H} = \mathbb{C}^2 \otimes \mathbb{C}^2$:
\begin{equation}
\begin{array}{c}
\begin{IEEEeqnarraybox}{lClClCl}
\ket{\beta_{00}} &=& \frac{\ket{00} + \ket{11}}{\sqrt{2}} &\quad&
\ket{\beta_{10}} &=& \frac{\ket{00} - \ket{11}}{\sqrt{2}} \\
\ket{\beta_{01}} &=& \frac{\ket{01} + \ket{10}}{\sqrt{2}} &&
\ket{\beta_{11}} &=& \frac{\ket{01} - \ket{10}}{\sqrt{2}}
\end{IEEEeqnarraybox}
\end{array}
\label{eq:bell-states}
\end{equation}
where $\ket{ij} = \ket{i} \otimes \ket{j}$ ($i,j \in \{0,1\}$) is the tensor product basis.
They are maximally entangled in the sense of entanglement entropy, which is a quantity defined for any pure bipartite state $\rho=\dyad{\psi}$ of two quantum systems $A$ and $B$ as
\begin{equation} 
E(\ket{\psi}) = S(\rho_A)/{\ln 2} = S(\rho_B)/{\ln 2}
\label{eq:entangentropy}
\end{equation}
where $S(\rho)=-\Tr{\rho\ln{\rho}}$ is the Von Neumann entropy and $\rho_{A/B}=\Tr_{A/B}(\rho)$ are the reduced density matrices of the two subsystems. As is well known Schmidt's theorem implies that their entropies are equal. A pure Bell state has maximal entropy of entanglement $1$ since its reduced density matrix is always $\frac{1}{2}\mathcal{I}$, $\mathcal{I}$ being the $2\times 2$ identity matrix, while a pure product state has vanishing entropy of entanglement.

By definition, the larger class of Bell-diagonal states (BDS) is the set of mixed states that are diagonal in the Bell basis,
i.e.~those given by density operators of the form
\begin{equation} 
\rho = \sum_{j,k = 0}^1 p_{jk} \dyad{\beta_{jk}}
\label{eq:bds}
\end{equation}
where $\{p_{jk}\}_{j,k = 0}^1$ is a set of probabilities summing to $1$.

Any two-qubit density matrix can be expanded on the basis of products of Pauli matrices $\sigma_j$, $j=1,2,3$ completed by the identity matrix $\mathcal{I}$
\begin{equation}
\rho = \frac{1}{4}\Bigg( \mathcal{I}\otimes\mathcal{I} + \pmb{r}\cdot\pmb{\sigma}\otimes\mathcal{I} + \mathcal{I}\otimes \pmb{s}\cdot\pmb{\sigma} + \sum^3_{n,m=1} t_{nm}\,\sigma_n\otimes\sigma_m
\Bigg)
\label{eq:rho_gen}
\end{equation} 
where $\pmb{r}\cdot\pmb{\sigma} = \sum^{3}_{i=1} r_i\sigma_i$ is the usual scalar product. Among the 15 real expansion parameters we find two vectors $\pmb{r}$ and $\pmb{s}$ in $\mathbb{R}^3$ corresponding to the marginal density matrices, and a $3\times 3$ pure correlation matrix $T$ (matrix elements $t_{nm}$). Thus states with maximally mixed marginals like BDS fulfill the conditions $\pmb{r} = 0$ and  $\pmb{s}=0$. Such states are, up to suitable local unitary transformations $U_A\otimes U_B$, equivalent to states with a diagonal $T$ matrix \cite{PhysRevA.54.1838}, moreover the latter states can always be considered as convex combinations of the four Bell-states \cite{PhysRevA.54.1838}, i.e. BDS. The  subset of all BDS density matrices ($p_{jk}\geq 0$) is thus a very interesting set which is fully characterized by a solid geometric tetrahedron $\mathcal{T}$ in the ``$t$-configuration'' space (see \cref{fig:BDS_geometry}), where each point $(t_{1},t_{2},t_{3})$ correspond to a density matrix with purely diagonal parameters $(t_1,t_2,t_3)$ \cite{PhysRevA.54.1838}
\begin{equation}
\rho = \frac{1}{4}\bigg( \mathcal{I}\otimes\mathcal{I} + \sum^3_{i=1} t_i\,\sigma_i\otimes\sigma_i
\Bigg)
\label{eq:rho_tuple}
\end{equation}
The importance of the tetrahedron geometry and the parameterization $\{p_{jk}\}_{j,k = 0}^1 \rightarrow (t_1,t_2,t_3)$ is going to be highlighted by the circuit analysis infra.
Let us write the formulas allowing to go from one representation to another:
\begin{equation}
\begin{array}{c}
\begin{IEEEeqnarraybox}{lClClCl}
p_{00} &=& \frac{1 + t_1 - t_2 + t_3}{4}
&\quad&
p_{10} &=& \frac{1 - t_1 + t_2 + t_3}{4}\\
p_{01} &=& \frac{1 + t_1 + t_2 - t_3}{4} &&
p_{11} &=& \frac{1 - t_1 - t_2 - t_3}{4}
\end{IEEEeqnarraybox}
\end{array}
\label{eq:param_prob}
\end{equation}
\begin{equation}
\begin{array}{c}
\begin{IEEEeqnarraybox}{lCl}
t_1 &=& p_{00}+p_{01}-p_{10}-p_{11} \\
t_2 &=& -p_{00}+p_{01}+p_{10}-p_{11} \\
t_3 &=& p_{00}-p_{01}+p_{10}-p_{11}
\end{IEEEeqnarraybox}
\end{array}
\label{eq:param_t}
\end{equation}
	
In \cref{sect:circuits}, we propose circuits that generate all BDS, i.e. in the entire tetrahedron $\mathcal{T}$. Moreover we characterize these states in terms of various quantum correlation and entanglement measures
(\cref{sect:entanglement_discord}) and see to what extent these can be tested on the NISQ devices of IBM Q (\cref{sect:IBMQ_results}). However, there are still two distinguished subsets of the tetrahedron which are interesting to discuss, namely the {\it octahedron of separable states} and the {\it line of Werner states}.  
	
\subsection{The octahedron of separable states}
	
As an appetizer let us first remark that the four corners of the tetrahedron are precisely the four Bell states that are maximally entangled in the sense that their entropy of entanglement is maximal. However the entropy of entanglement {\it cannot} be defined for mixed states. Indeed any tensor product of two genuinely mixed states $\rho_A\otimes \rho_B$ has a reduced density matrix with possibly different reduced von Neumann entropies. To give meaningful measures of ``entanglement'' and other ``quantum correlation'' for mixed states it is necessary to generalize the notion of product pure states. For bipartite systems, {\it separable} mixed states are usually defined as an arbitrary convex superposition of products of density matrices
\begin{align}\label{eq:sep}
 \rho = \sum_i q_i \rho_A^{(i)} \otimes \rho_B^{(i)}, \quad q_i\in [0, 1], \quad \sum_i q_i=1
\end{align}
Non-separable mixed states are the ones that cannot be represented as such, and are called {\it entangled}. It is clear that for a separable state the partial transpose $\rho_B^{T_B}$ must necessarily admit only non-negative eigenvalues (indeed $\rho_B^{(i)}$ is positive semidefinite, thus $\rho_B^{(i) T}$ also is, and therefore \eqref{eq:sep} implies that $\rho^{T_B}$ must be positive semidefinite). This is the so-called Positive Partial Transpose (PPT) criterion of Peres
\cite{Peres1996}. Remarkably, for $2\otimes 2$ and $2\otimes 3$ bipartite density matrices $\rho_{AB}$ the PPT criterion is necessary and sufficient \cite{HORODECKI19961}. We refer to \cite{QEntanglementReview} for more details. 
	
\begin{figure}[h]
	\centering
	\includegraphics[width=0.3\textwidth]{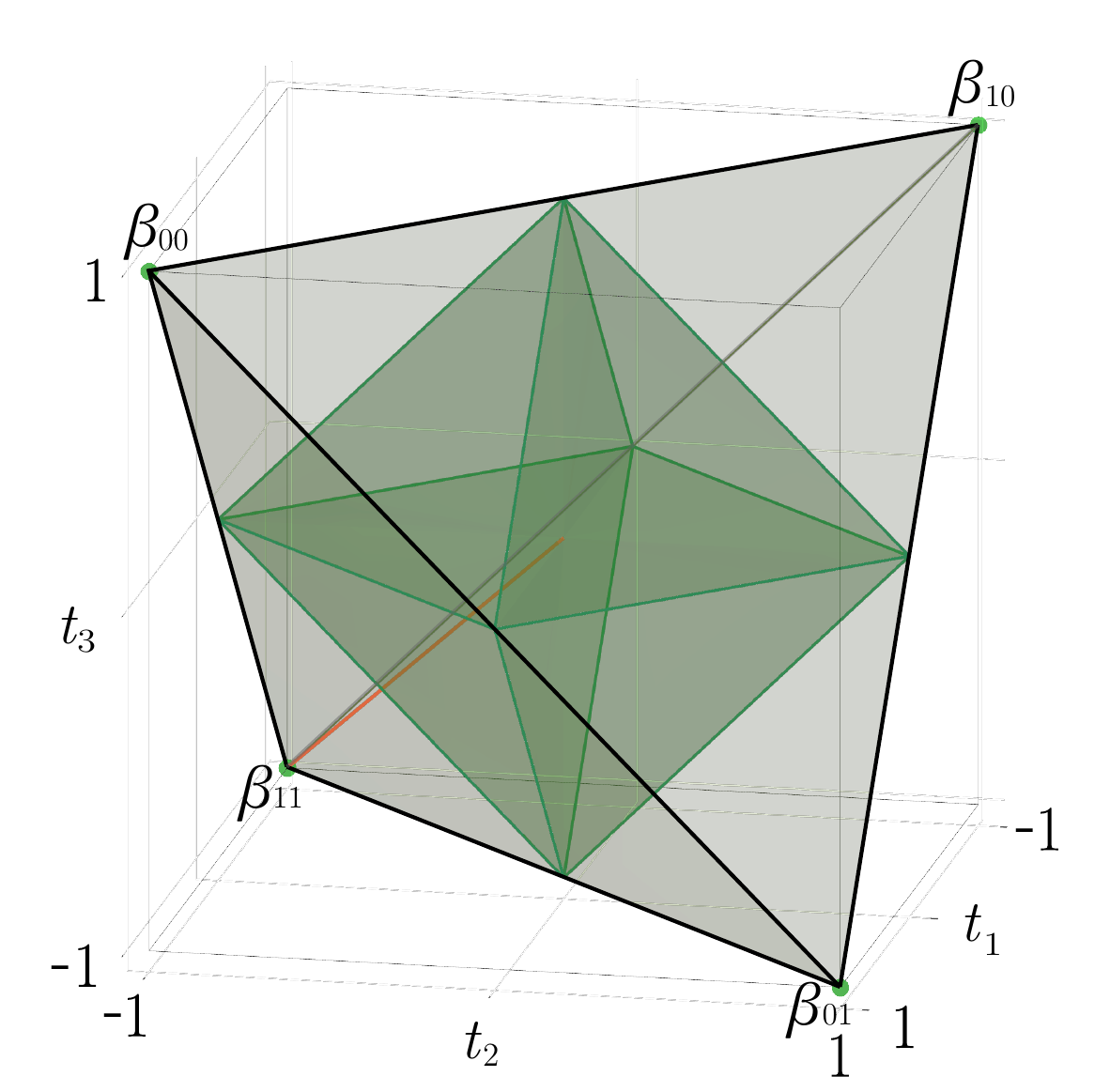}
	\caption{{\small Geometrical representation of the BDS tetrahedron $\mathcal{T}$ bounded by the four planes 
	$t_1-t_2+t_3\geq -1$, $t_1+t_2-t_3\geq -1$, $t_1 -t_2-t_3 \leq 1$, $t_1+t_2+t_3\leq 1$. The octahedron $\mathcal{O}$ defined by $\vert t_1\vert +\vert t_2\vert + \vert t_3\vert \leq 1$ contains all the separable BDS. Accordingly, four entangled regions can be identified outside of the octahedron, in each of which a Bell state is located at the corresponding summits of $\mathcal{T}$. We have the correspondence 
	$\vert\beta_{00}\rangle \leftrightarrow (1, -1, 1)$, $\vert\beta_{01}\rangle \leftrightarrow (1,1,-1)$, $\vert\beta_{10}\rangle\leftrightarrow (-1,1,1)$, $\vert\beta_{11}\rangle \leftrightarrow (-1,-1,-1)$. The red line $t_1=t_2=t_3=-w$, $0\leq w\leq 1$, along the negative diagonal, represents Werner states \eqref{eq:rho_werner}.}}
	\label{fig:BDS_geometry}
\end{figure}

In the case of BDS we easily see from 
\eqref{eq:rho_tuple} that the partial transpose of a BDS parameterized by $(t_1, t_2, t_3)$ is a matrix parameterized by $(t_1, -t_2, t_3)$ (because $\sigma_1^T=\sigma_1$, $\sigma_2^T=-\sigma_2$, $\sigma_3^T=\sigma_3$). Thus, partially transposed BDS correspond to a reflected tetrahedron obtained from $\mathcal{T}$ by a reflection across the $(t_1, t_3)$ plane. The intersection of this reflected tetrahedron with $\mathcal{T}$ is an octahedron $\mathcal{O}= \{(t_1, t_2, t_3)\mid \vert t_1\vert + \vert t_2\vert + \vert t_3\vert \leq 1\}$ (\cref{fig:BDS_geometry}). All elements of $\mathcal{O}$ must correspond to bona-fide density matrices with non-negative eigenvalues. Therefore points of $\mathcal{O}$ necessarily correspond to separable mixed states. We still must check that points of $\mathcal{T}\setminus \mathcal{O}$ correspond to non-separable states.  First note that under the reflection across the $(t_1,t_3)$ plane these points are sent outside of $\mathcal{T}$. By the PPT criterion it suffices to see that such a point corresponds to a matrix with at least one negative eigenvalue. This last claim is checked by contradiction. Indeed the (partially transposed) matrix has trace one, so, if  all its eigenvalues were {\it non-negative}, they would also be smaller or equal to one, hence the matrix would be a density matrix of the form \eqref{eq:rho_tuple}, hence a BDS belonging to $\mathcal{T}$, a contradiction.
	
Finally, let us note that the Bell states $\ket{\beta_{ij}}$  are the ``furthest apart'' from the subset of separable BDS, confirming that Bell states are maximally entangled.
	
The PPT criterion as such is only qualitative and discriminates efficiently separable and entangled mixed states. However it should be pointed out that the corresponding amount of negativity, defined as
\begin{equation}
{\cal N}(\rho) = \frac{1}{2} \left(\Vert\rho_B^{T_B}\Vert_1-1\right)
\label{eq:negativity}
\end{equation}
is quantitative in the sense that it is an entanglement monotone (here $\Vert.\Vert_1$ is the trace norm). Nevertheless it does not address the question of a measure of ``quantumness'' of correlations other than entanglement. This issue is discussed in \cref{sect:entanglement_discord}.

\subsection{The line of Werner states}

A particularly interesting subset of BDS is formed by Werner states \cite{PhysRevA.40.4277}, which for 2 qubits are defined by the parameter $w=-t_1=-t_2=-t_3$:
\begin{equation}
\rho = \frac{(1-w)}{4} \mathcal{I}\otimes\mathcal{I} + w\,\ket{\beta_{11}}\bra{\beta_{11}}
\label{eq:rho_werner}
\end{equation}
Geometrically, they are represented by a straight line inside the BDS tetrahedron (red line in \cref{fig:BDS_geometry}). On the one side, the $w=0$-extremity of the segment corresponds to the state $\rho= \frac{1}{4}\sum_{j,k=0}^1 \dyad{\beta_{jk}}$, which is a uniform statistical mixture of all Bell states. The $w=1$-extremity refers to the maximally entangled state $\ket{\beta_{11}}$. More generally, the PPT criterion applies and shows that Werner states are separable for $w\in [0, 1/3]$  and entangled for $w\in (1/3, 1]$. The critical value $w=1/3$ corresponds exactly to the intersection of the red line in \cref{fig:BDS_geometry} with a face of the octahedron.

\section{Quantum circuits for BDS and Werner states}

In this section, we propose quantum circuits with output states covering the whole tetrahedron of BDS. We propose various circuits, using four qubits, as well as two qubits, and discuss their relationship with various parameterizations. Specialized circuits for Werner states are also considered. Some of these circuits serve as the basis for our implementation of BDS and their characterization on the IBM Q devices.
	
\label{sect:circuits}
\begin{figure}[h]
	\centering
	\[ \Qcircuit @C=1em @R=1em {
		\lstick{\ket{0}_a} & \multigate{1}{G} \barrier[-0.7em]{1}
		\save [0,1]+<-.7em,2em> *{\ket{\psi}_{ab}} \restore
		& \ctrl{2} & \qw & \qw & \qw & \qw \\
		\lstick{\ket{0}_b} & \ghost{G} & \qw & \ctrl{2} & \qw & \qw & \qw \\
		\lstick{\ket{0}_c} & \qw & \targ & \qw & \gate{H} & \ctrl{1} & \qw \\
		\lstick{\ket{0}_d} & \qw & \qw & \targ & \qw & \targ & \qw
		\save "4,5"+<1em,-1.5em> *{B} \restore
		\gategroup{3}{5}{4}{6}{.7em}{--}
		\outputgroupv{3}{4}{7}{.6em}{1.2em}{\hspace{-.8em}\rho_{cd}}
		\outputgroupv{1}{2}{7}{.6em}{1.1em}{\hspace{.2em}
    	\scriptsize\eqbox{c}{t}{unread\\(env.)}}
	} \]
	\caption{{\small The generalized four-qubit preparation circuit as in ref. \cite{Pozzobom2019}. Only the subcircuit $G$ which encodes the probabilities $\{p_{jk}\}_{j,k = 0}^1$ must be corrected. Qbits are then copied by CNOT gates. Subcircuit $B$ finally entangles into the Bell basis.}}
	\label{fig:four-qubit}
\end{figure}
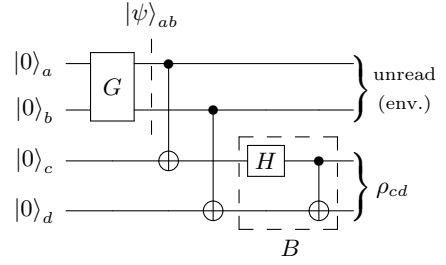
\begin{figure}[h]
	\centering
	\begin{subfigure}{.49\textwidth}
		\[ \Qcircuit @C=1em @R=1em {
			\lstick{\ket{0}_a} & \gate{R_y(\theta)} & \qw \\
			\lstick{\ket{0}_b} & \gate{R_y(\alpha)} & \qw
			\save "1,2"+<0em,2em> *{G} \restore
			\gategroup{1}{2}{2}{2}{.7em}{--}
			\outputgroupv{1}{2}{3}{.6em}{1.3em}{\ket{\psi}_{ab}}
		} \]
		\caption{{\small Encoder $G$: incomplete circuit of \cite{Pozzobom2019}, which involves only two parameters $\alpha$ and $\theta$ (definition in \cite{Pozzobom2019}), and the gate $R_y$ given by eq. \eqref{eq:Ry-rot}.}}
		\label{fig:G-pozzobom}
	\end{subfigure}
	\begin{subfigure}{.49\textwidth}
		\[ \Qcircuit @C=1em @R=1em {
			\lstick{\ket{0}_a} & \gate{R_y(\alpha)} & \ctrl{1} & \gate{R_y(\beta)} & \qw \\
			\lstick{\ket{0}_b} & \qw & \targ & \gate{R_y(\gamma)} & \qw
			\save "1,3"+<0em,2em> *{G} \restore
			\gategroup{1}{2}{2}{4}{.7em}{--}
			\outputgroupv{1}{2}{5}{.6em}{1.3em}{\ket{\psi}_{ab}}
		} \]
		\caption{{\small Encoder $G$: compact circuit which generates the whole class of BDS with the three parameters $\alpha$, $\beta$ and $\gamma$ appearing in eq. \eqref{eq:compact-probs}.}}
		\label{fig:G-compact}
	\end{subfigure}
	\begin{subfigure}{.49\textwidth}
		\[ \Qcircuit @C=1em @R=1em {
			\lstick{\ket{0}_a} & \qw & \qw & \gate{R_y(2\theta)} & \ctrl{1} & \qw & \qw \\
				\lstick{\ket{0}_b} & \qw & \gate{R_y(2\psi)} & \ctrl{-1} & \gate{R_y(-2\varphi)} &
				\qw & \qw
			\save "1,4"+<0em,2em> *{G} \restore
			\gategroup{1}{3}{2}{5}{1.9em}{--}
			\outputgroupv{1}{2}{7}{.6em}{1.3em}{\ket{\psi}_{ab}}
		} \]
		\caption{{\small Encoder $G$: complete three-parameter circuit based on canonical coordinates $\psi$, $\theta$ and $\varphi$ on the unit 3-sphere appearing in eq. \eqref{eq:canonical-probs}.}}
		\label{fig:G-canonical}
	\end{subfigure}
	\caption{{\small Three different versions of the probability-encoding subcircuit $G$.}}
	\label{fig:G-variants}
\end{figure}
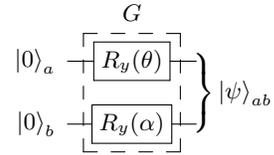
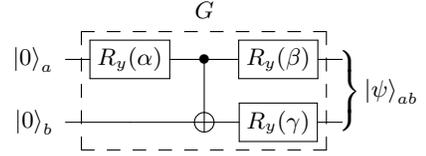
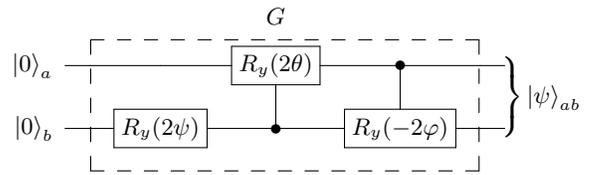
	
\subsection{Four qubit circuits and relevant BDS parameterizations}
	
Following Pozzobom and Maziero \cite{Pozzobom2019}, we consider a four-qubit circuit of the form portrayed in \cref{fig:four-qubit}. The subcircuit $G$ is tasked with encoding the probabilities $\{p_{jk}\}$ in a two-qubit state
\begin{equation}
G\ket{00} \equiv \ket{\psi} \equiv \sum_{j,k=0}^1 \sqrt{p_{jk}} \ket{jk}
\label{eq:psi}
\end{equation}
This is mapped to
\begin{equation}
\sum_{j,k=0}^1 \sqrt{p_{jk}} \ket{jk}_{ab} \otimes \ket{jk}_{cd}
\label{eq:psi-four-qubit}
\end{equation}
by the two controlled-NOT (CNOT) gates. Finally, the Bell basis change transformation $B$ is applied.
Note that we swapped the qubits in $B$ (w.r.t. \cite{Pozzobom2019}), to fit standard Bell state conventions. It produces $B\ket{jk} = \ket{\beta_{jk}}$, so that the resulting state is
\begin{equation}
\ket{\tau} \equiv \sum_{j,k=0}^1 \sqrt{p_{jk}} \ket{jk}_{ab} \otimes \ket{\beta_{jk}}_{cd}.
\end{equation}
This is a purification of the BDS $\rho$ from eq. \eqref{eq:bds}, meaning that one can retrieve $\rho$
by considering the first two qubits as part of the environment, which amounts to a partial trace
operation:
\begin{equation}
\Tr_{ab}(\dyad{\tau}) = \rho.
\end{equation}
	
Now, we turn to the probability encoder $G$. Pozzobom and Maziero used the two-parameter subcircuit shown in \cref{fig:G-pozzobom}. There the $y$-rotation gate is given by
\begin{equation}
    R_y(\theta) = e^{-i\frac{Y}{2} \theta} =
    {\footnotesize\left(\begin{array}{c}
          \begin{IEEEeqnarraybox}{rCr}
          \cos(\theta/2) &\:& -\sin(\theta/2) \\
          \sin(\theta/2) && \cos(\theta/2)
          \end{IEEEeqnarraybox}
     \end{array} \right)}
\label{eq:Ry-rot}
\end{equation}
in the computational basis.
But, as we have already noted, two parameters are not enough to cover all choices of $\{p_{jk}\}$; in fact, one solely gets those that can be factored as
$p_{jk} = a_j b_k$ for some $\{a_j\}$ and $\{b_k\}$.
This is a direct consequence of the failure of their encoder $G$ to entangle the two qubits $a$ and $b$.
	
A better working implementation of $G$ is displayed in \cref{fig:G-compact}. It is perhaps the simplest conceivable implementation: it cannot be simplified to use less than three parameters, and it does entangle the two qubits $a$ and $b$; both of these are necessary features of any working encoder. The output state is given by \eqref{eq:psi} with the probabilities
\begin{equation}
\begin{array}{c}
  {\footnotesize
	\begin{IEEEeqnarraybox}{lCl}
	\sqrt{p_{00}} &=& \hcos{\alpha} \hcos{\beta} \hcos{\gamma}
	+ \hsin{\alpha} \hsin{\beta} \hsin{\gamma} \\
	\sqrt{p_{01}} &=& \hcos{\alpha} \hcos{\beta} \hsin{\gamma}
	- \hsin{\alpha} \hsin{\beta} \hcos{\gamma} \\
	\sqrt{p_{10}} &=& \hcos{\alpha} \hsin{\beta} \hcos{\gamma}
	- \hsin{\alpha} \hcos{\beta} \hsin{\gamma} \\
	\sqrt{p_{11}} &=& \hcos{\alpha} \hsin{\beta} \hsin{\gamma}
	+ \hsin{\alpha} \hcos{\beta} \hcos{\gamma}.\vspace{0.5mm}
	\end{IEEEeqnarraybox}
  }
\end{array}\vspace{1mm}
\label{eq:compact-probs}
\end{equation}
In order to prepare any given Bell-diagonal state, one thus writes it in the form of eq. \eqref{eq:bds} and solves \eqref{eq:compact-probs} to obtain the corresponding parameters $\alpha$, $\beta$ and $\gamma$. It is straightforward to solve eqs.~\eqref{eq:compact-probs} numerically. An analytical solution exists as well and is given in Appendix~\ref{app:analytical-solution}.
	
An alternative realization of $G$ is displayed in \cref{fig:G-canonical}. It uses two controlled
$y$-rotation gates, e.g.~$C_{R_y(2\varphi)}^{a \to b} = \dyad{0}_a \otimes \mathcal{I}_b
+ \dyad{1}_a \otimes R_y(2\varphi)_b$. This circuit realizes the canonical hypersphere coordinates
\begin{equation}
\begin{array}{c}
\begin{IEEEeqnarraybox}{lCl}
\sqrt{p_{00}} &=& \cos(\psi) \\
\sqrt{p_{01}} &=& \sin(\psi) \cos(\theta) \\
\sqrt{p_{11}} &=& \sin(\psi) \sin(\theta) \cos(\varphi) \\
\sqrt{p_{10}} &=& \sin(\psi) \sin(\theta) \sin(\varphi)
\end{IEEEeqnarraybox}
\end{array}
\label{eq:canonical-probs}
\end{equation}
(note the ordering $00$--$01$--$11$--$10$: the two-bit Gray code) which have the advantage of being easily obtainable in terms of $\{p_{jk}\}$ by calculating their cosines in an iterative manner, as follows:
\begin{equation}
\begin{array}{c}
\begin{IEEEeqnarraybox}{lCl}
\cos^2(\psi) &=& p_{00} \\
\cos^2(\theta) &=& \frac{p_{01}}{1 - \cos^2(\psi)} \\
\cos^2(\varphi) &=&
\frac{p_{11}}{\qty(1 - \cos^2(\psi))\qty(1 - \cos^2(\theta))}.
\end{IEEEeqnarraybox}
\end{array}
\label{eq:canonical-probs-inverse}
\end{equation}
When evaluating these expressions, any quotient $0/0$ is taken to be $1$ (in practice, one must also beware of rounding errors). The circuit in question is more transparent than the one suggested supra. Nevertheless, its circuit complexity is higher, since each controlled rotation will typically be implemented using two CNOT gates as well as several one-qubit unitaries. In the sequel, we will provide an operational comparison of the two circuits to assess how severe this problem is.
	
\subsection{Two qubit circuits}
\label{sec:two-qubit}
	
\begin{figure}[h]
	\[ \Qcircuit @C=1em @R=1em {
		\lstick{\ket{0}_a} & \multigate{1}{G} \barrier[-1.35em]{1}
		\save [0,1]+<-1.35em,2em> *{\ket{\psi}_{ab}} \restore
		& \meter \barrier[-0.2em]{1}
		\save [0,1]+<-0.2em,2em> *{R_{ab}} \restore
		& \qw & \gate{H} & \ctrl{1} & \qw \\
		\lstick{\ket{0}_b} & \ghost{G} & \meter & \qw & \qw & \targ & \qw \\
		%
		\save "1,5"+<1em,1.7em> *{B} \restore
		\gategroup{1}{5}{2}{6}{.7em}{--}
		\outputgroupv{1}{2}{7}{.6em}{1.3em}{\hspace{-.8em}\rho_{ab}}
	} \]
	\caption{{\small A two-qubit replacement for the circuit in \cref{fig:four-qubit}.}}
	\label{fig:two-qubit}
\end{figure}
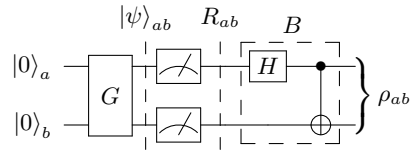

The circuit template in \cref{fig:four-qubit} uses four qubits, which seems inefficient as the objective is to prepare a two-qubit output state. In fact, one could remove the two ancillary qubits and instead perform unread measurements on the principal qubits, as shown in \cref{fig:two-qubit}. The measurements collapse the pure state $\ket{\psi}$, as given in \eqref{eq:psi}, into the mixture
\begin{equation}
R \equiv \sum_{j,k=0}^1 p_{jk} \dyad{jk}
\end{equation}
of computational basis states. Thus the combination of $G$ and the measurements acts as a ``quantum random number generator''. Finally, applying $B$ transforms $R$ into the prescribed Bell-diagonal state \eqref{eq:bds}:
\begin{equation}
BRB^\dag = \sum_{j,k=0}^1 p_{jk} \dyad{\beta_{jk}} = \rho.
\end{equation}
The two-qubit circuit works through unread measurements, which can be interpreted as unmonitored interactions with the environment. 

\begin{figure}[h]
	\centering
	\[ \Qcircuit @C=1em @R=1em {
		\lstick{\ket{0}_a} & \multigate{1}{G} \barrier[-0.7em]{1}
		\save [0,1]+<-.7em,2em> *{\ket{\psi}_{ab}} \restore
		& \ctrl{2} \barrier[0.5em]{1}
		\save [0,1]+<0.5em,2em> *{R_{ab}} \restore & \qw & \gate{H} & \ctrl{1} & \qw \\
		\lstick{\ket{0}_b} & \ghost{G} & \qw & \ctrl{2} & \qw & \targ & \qw \\
		\lstick{\ket{0}_c} & \qw & \targ & \qw & \qw & \qw & \qw \\
		\lstick{\ket{0}_d} & \qw & \qw & \targ & \qw & \qw & \qw
		\save "1,5"+<1em,1.7em> *{B} \restore
		\gategroup{1}{5}{2}{6}{.7em}{--}
		\outputgroupv{1}{2}{7}{.6em}{1.2em}{\hspace{-.8em}\rho_{ab}}
		\outputgroupv{3}{4}{7}{.6em}{1.1em}{\hspace{.2em}
    	\scriptsize\eqbox{c}{t}{unread\\(env.)}}
	} \]
	\caption{{\small A four-qubit circuit illustrating measurement as entanglement with the environment.}}
	\label{fig:four-qubit-ancillary}
\end{figure}
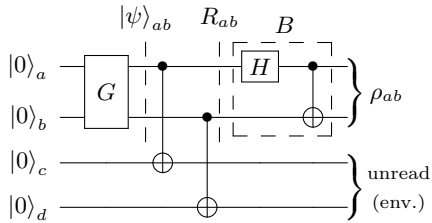

\Cref{fig:two-qubit,fig:four-qubit-ancillary} jointly illustrate this equivalence: one can implement an unread measurement of a system as a unitary evolution of that system together with an environment. Moreover, it is evident from the symmetry of \cref{eq:psi-four-qubit} that \cref{fig:four-qubit-ancillary} describes an equivalent circuit to that of \cref{fig:four-qubit}. \Cref{fig:four-qubit-ancillary} thus shows the connection between the two-qubit and four-qubit versions.
	
Out of the three equivalent circuits described supra., (\cref{fig:four-qubit,fig:two-qubit,fig:four-qubit-ancillary}), we suggest the two-qubit variant. Using four qubits amounts to utilizing precious resources to simulate decoherence on a coherent system rather than making use of decoherence already redundantly available in today's noisy quantum computers.
	
\begin{figure}[h]
	\centering
	\[ \Qcircuit @C=1em @R=1em {
		\lstick{\ket{0}_a} & \qw & \gate{R_y(\theta)} & \meter \cwx[2] & \qw & \gate{H} & \ctrl{1}
		& \meter \cwx[1] & \qw \\
		\lstick{\ket{0}_b} & \qw & \qw & \qw & \gate{X} & \gate{H} & \targ & \meter & \qw \\
		\lstick{0_c} & \cw & \cw & \cw & \cctrl{-1} & \cctrl{-1} & \cctrl{-1} & \cctrl{-1}
		\save "3,5"+<0em,-.7em> *{\scriptstyle c = 1} \restore
		\save "3,6"+<0em,-.7em> *{\scriptstyle c = 0} \restore
		\save "3,7"+<0em,-.7em> *{\scriptstyle c = 1} \restore
		\save "3,8"+<0em,-.7em> *{\scriptstyle c = 0} \restore
		\outputgroupv{1}{2}{9}{.6em}{1.3em}{\hspace{-.8em}\rho_{ab}}
	} \]
	\caption{{\small A specialized circuit for the preparation of Werner states.}}
	\label{fig:werner-circuit}
\end{figure}
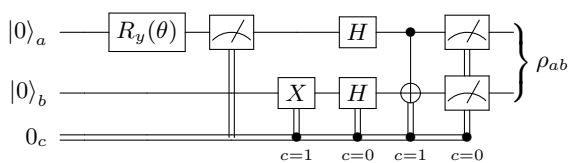

The circuits proposed above can prepare Werner states, as the latter form a subset of the Bell-diagonal states. Nevertheless, in cases where the full range of BDS is not needed, specializing the circuit offers opportunities for a further optimization. The circuit shown in \cref{fig:werner-circuit} prepares the Werner state given in \eqref{eq:rho_werner} using classically controlled quantum operations. First, qubit $a$ is put into a superposition $\sqrt{1 - w} \ket{0} + \sqrt{w} \ket{1}$, where we selected the parameter $\theta$ such that $\sqrt{w} = \sin(\frac{\theta}{2})$.
Then, the state is measured, giving $1$ with probability $w$ and $0$ with probability $1 - w$, storing the outcome in a classical bit $c$. This first part is, again, a quantum random number generator. If the outcome is $0$, the circuit prepares the maximally mixed state $\frac{1}{4}\mathcal{I} \otimes \mathcal{I}$ by generating $\ket{++} \equiv H^{\otimes 2} \ket{00}$ and performing an unread measurement. If the outcome is $1$, it prepares the pure state $\ket{\beta_{11}} = B\ket{11}$ by flipping the lower qubit to $\ket{1}$ (the upper one is already $\ket{1}$) and applying the Bell basis change $B$.

All two-qubit circuits proposed so far rely on applying further quantum gates after performing a measurement on a qubit. Such operations are not supported on present-day IBM devices, and as a result, only the four-qubit circuits may be run on real hardware. However, the two-qubit variants can be simulated in Qiskit, as detailed in the next section. 

The last two-qubit circuit of \cref{fig:werner-circuit} incorporates in addition parallel classical information treatment and classically controlled quantum gates. This too is not yet possible with current hardware, but we show in Appendix~\ref{app:Q-implementation-classical-op} how it could be replaced by an equivalent fully quantum circuit. There we see that the number of necessary qubits would rise to five, and many more quantum gates and computational steps would be required, making probably such alternatives largely unattractive because of enhanced decoherence.

\section{Qiskit implementation}\label{sec:Qiskitimplementation}
To run the quantum circuits described in \cref{sect:circuits} on IBM Q hardware, we have provided
implementations using Qiskit \cite{Qiskit}, available in a Git repository hosted on GitLab
\cite{gitlab-repo}. The software performs several functions.

The basic functionality is circuit construction. For a choice of probability encoder from 
\cref{fig:G-variants} and a four-qubit or two-qubit template (\cref{fig:four-qubit} or 
\cref{fig:two-qubit}), and given the parameters $\{p_{jk}\}_{j,k=0}^1$,
the software constructs a Qiskit representation of the quantum circuit for preparing the
corresponding BDS, computing the correct circuit parameters such as $(\alpha,\beta,\gamma)$
in the process. The specialized circuit from \cref{fig:werner-circuit} can also be constructed
for any given $w$.

The software also performs quantum state tomography \cite{tomography-tutorial} to reconstruct the
output state of the circuits with a new set of routines. Qiskit has built-in routines for tomography, but they require some amendment for use on circuits that contain classical registers, including
\cref{fig:two-qubit,fig:werner-circuit} (implementing
\cref{fig:two-qubit} in Qiskit does require classical registers as destinations for the
unread measurement results, although they are implicit in the figure).
Specifically, the built-in routines determine the output state of an $n$-qubit circuit
by performing various operations
indexed by $k$, on the output and then performing a
measurement into an added $n$-bit classical register $a_1 \dots a_n$.
Each type of measurement $k$ is repeated multiple
times (\textit{shots}), and the results are presented as \textit{counts} $n^k_{b_1 \dots b_n}$,
where $b_i \in \{0,1\}$, giving the number of times the measured bit string was
$a_1 \dots a_n = b_1 \dots b_n$. If the original circuit had $m$ classical registers
$c_1 \dots c_m$, the resulting counts $n^k_{b_1 \dots b_n d_1 \dots d_m}$ need to
be aggregated as
\[ n^k_{b_1 \dots b_n} \equiv \sum_{d_1 = 0}^1 \dots \sum_{d_m = 0}^1
    n^k_{b_1 \dots b_n d_1 \dots d_m} \]
before being passed to the built-in tomographic reconstruction routine.
The implementation partly follows \cite{tomography-tutorial}.
See also \cite{github-issue-tomography}.

A further complication arises when implementing the specialized circuit for Werner states,
\cref{fig:werner-circuit}. This circuit contains a measurement operation conditioned on
the value of a classical bit. Such conditional measurements are not officially supported
by the Qiskit Aer simulator. However, the circuit can be simulated using a custom version
of Qiskit Aer where a small change has been made to the C++ source before compiling
\cite{github-issue-conditional-measurements}.
Our GitLab repository \cite{gitlab-repo} contains a patch with the necessary changes.

The implementations have been tested for correctness. All four combinations of
\cref{fig:G-compact} or \cref{fig:G-canonical} together with \cref{fig:four-qubit} or
\cref{fig:two-qubit} were simulated in Qiskit, without noise, for 340 states uniformly
distributed in the BDS tetrahedron.
The circuit of \cref{fig:werner-circuit} was also run on 100 Werner states, uniformly distributed
between $w = 0$ and $w = 1$.
The density matrices of the output states were reconstructed via tomography, as detailed above,
with $2^{10}$ shots each, and the state fidelity \eqref{eq:fid} was computed.
For all circuits, the mean fidelity was 99.5\% with a standard deviation of 0.5\%.

\section{Entanglement measures and discord}\label{sect:entanglement_discord}

In this section, we shall review entanglement and correlation measures for BDS. These fall in three categories: non-separability (entanglement), non-locality and steering measures. However going through fundamental operational definitions of all these notions would go out of the scope of this paper (see e.g. \cite{Wiseman-et-al2007} and \cite{Jones-et-al2007}). In \cref{sec:entanglementmeasures}, we focus on main known criteria that allow specific closed form formulas for BDS: entanglement of formation and concurrence, a restricted setting of CHSH-non-locality implied by Bell non-locality, and a highly restricted form of steering for which BDS are a useful testing ground. In \cref{sec:discord},  we further develop the notion of discord which quantifies the non-classical (i.e. quantum) correlations that are not necessarily related to entanglement. The relationship between original discord and asymmetric relative entropy of discord is profoundly reexamined and we show a new general inequality between the two quantities, and prove that for BDS they are equal. Specific expressions are computed as a function of $(t_1, t_2, t_3)$ whenever possible, and their behavior in the whole tetrahedron is illustrated. Finally \cref{sec:wernermeasures} focuses on the one-parameter family of Werner states which forms a very interesting particular special case. The theoretical results summarized here will serve as benchmarks for the quality of BDS and Werner states created by our circuits on IBM Q.

\subsection{Entanglement  measures for BDS}\label{sec:entanglementmeasures}
\subsubsection{Entanglement of formation and concurrence}
Entanglement of formation is the first metric of entanglement which properly extends to mixed states the notion of entanglement entropy $E(\psi)$ introduced in \cref{sect:BDS_WS_intro}. Strictly speaking the entanglement of formation $E_F(\rho)$ of a mixed state $\rho$ is the minimum average entanglement entropy over any ensemble of pure states that would represent the mixed state $\rho= \sum_i p_i \ket{\psi_i}\bra{\psi_i}$ (convex roof extension), and is defined as \cite{Bennett-et-al1996}
\begin{align}\label{eq:EoF}
E_F(\rho)=\min_{p_i, \ket{\psi_i}}{\sum_i p_i E(\psi_i)}
\end{align}
For a pure state entanglement of formation reduces to the entanglement entropy.

To understand it better it is useful to recall its remarkable operational meaning. Suppose two distant parties (Alice and Bob) share a large amount of Bell pairs $\vert\beta_{11}\rangle$ and suppose they want to convert them into roughly $n$ copies of $\vert \Phi_{AB}\rangle$, by using only LOCC. Then $n S(\rho_A)$ is roughly the minimum number of shared Bell pairs they need to ``burn'' (or spend) for this operation. If one views the $\ket{\beta_{11}}$ as a basic unit of entanglement, the ``ebit'', this means for example that a pure bipartite state with $S(\rho_A) = 1/10$ is ``equivalent'' (in LOCC sense) to one-tenth of an ebit. 

Now, let $\rho_{AB}$ a bipartite mixed state. it is possible to show that (asymptotically for $n\to +\infty$) 
the minimum number of Bell pairs needed by Alice and Bob to fabricate $n$ copies of $\rho_{AB}$ by using only LOCC is roughly $n E_F(\rho_{AB})$. This remarkable result was first derived by Bennett et al.~\cite{Bennett-et-al1996}.

Computing eq. \eqref{eq:EoF} is a difficult optimization problem. Happily, for arbitrary 2-qubit systems, Wootters~\cite{PhysRevLett.80.2245} derived a non-trivial closed form formula in terms of the {\it concurrence}. Let 
\begin{align}\label{eq:conc}
C(\rho)=\max\{{0, \mu_1 - \mu_2 - \mu_3 - \mu_4}\}
\end{align}
where $\mu_1 \geq \mu_2\geq \mu_3\geq\mu_4$ are the square roots of the four eigenvalues, in descending order, of the non-Hermitian matrix $\rho\tilde\rho$ where
$\tilde \rho = \sigma_2\otimes\sigma_2 \rho^* \sigma_2\otimes \sigma_2$ and $\rho^*$ the complex conjugated matrix in the computational basis representation. Then
\begin{align}\label{eq:wootters}
E_{F}(\rho) = h_2\Big(\frac{1}{2}(1 + \sqrt{1 - C(\rho)^2})\Big)
\end{align}
where $h_2(x) = -x \log_2 x - (1-x) \log_2(1-x)$ is the binary entropy function (with a range in $[0, 1]$ since one uses the log in base two)

For separable mixed states it easy to see that the entanglement of formation vanishes, to this end just insert the spectral decompositions of the factors in \eqref{eq:sep} and compute the corresponding sum of entanglement entropies. 

\begin{figure}[h]
		\centering
		\includegraphics[width=0.3\textwidth, trim = {2cm, 3cm, 2cm, 3cm}]{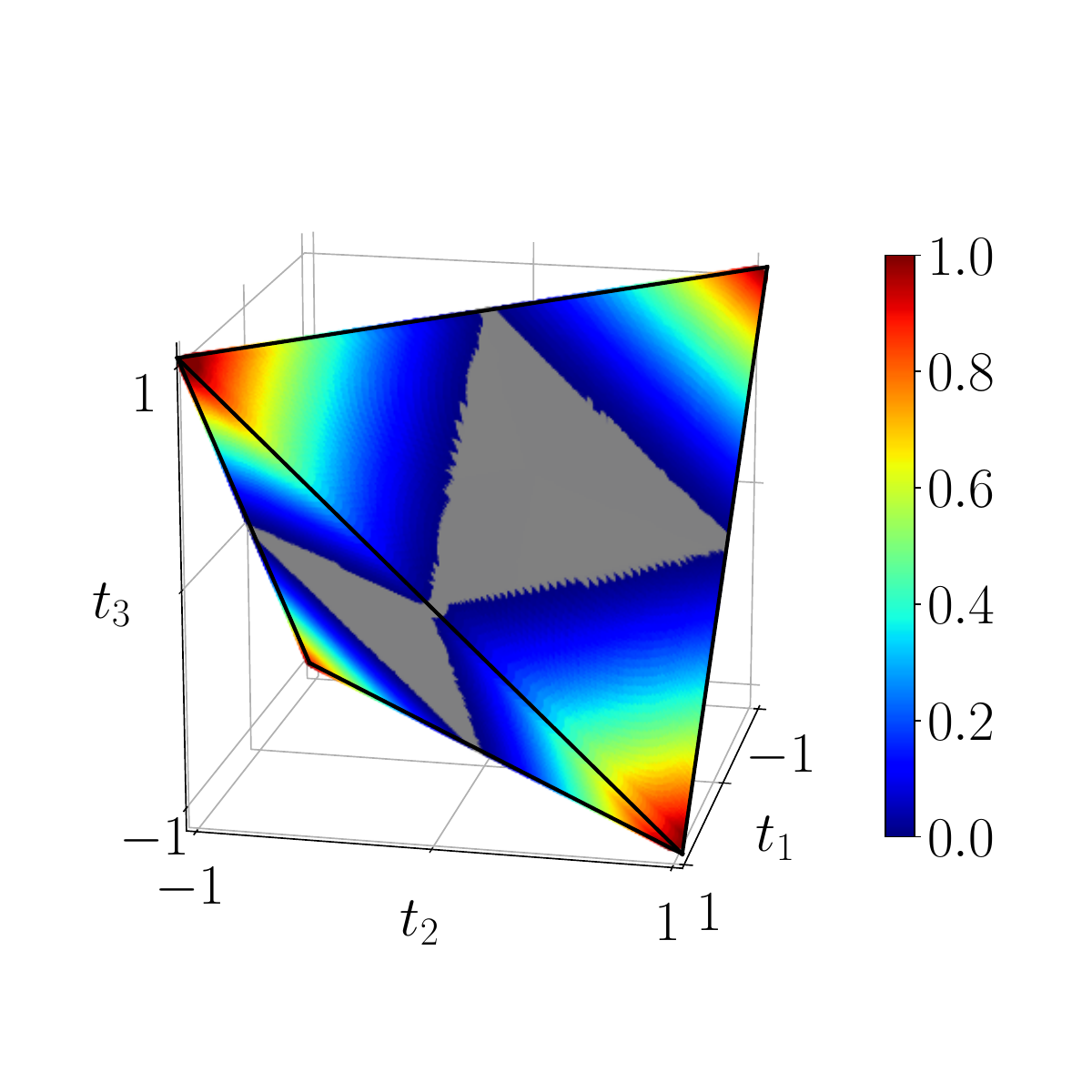}
		\caption{{\small Entanglement of formation $E_{F}(\rho)$ of BDS, calculated from the noiseless simulation of compact circuit  (\cref{fig:G-compact}). In the gray region entanglement of formation identically vanishes.}}
		\label{fig:EoF_2}
\end{figure}

\Cref{fig:EoF_2} displays the entanglement of formation $E_{F}(\rho)$ for all BDS in the tetrahedron, computed using the simulation circuit of \cref{fig:G-compact}. It can be shown that it corresponds exactly to the analytical result \eqref{eq:wootters}. We also see that entanglement of formation vanishes on the portion of the faces which are also faces of the octahedron of separable states. On the other hand for the four extremal Bell states entanglement of formation is maximal and equal to $1$ as expected.

The entanglement of formation on the interesting Werner line corresponding to Werner states (inner diagonal) will be better displayed in \cref{sec:wernermeasures} when we illustrate the corresponding explicit formula.

\subsubsection{CHSH-non-locality}
The fundamental definition of non-locality (or Bell-non-locality) expresses the fact that there is no local-hidden-variable (LHV) model allowing to explain all experimental joint histograms obtained by local measurements of two parties of a bipartite system. Suppose one wants to assess what is the part of the Hilbert space that displays non-locality. This is a priori difficult since all possible local measurements have to be examined. For this reason one reverts to criteria that give sufficient conditions for non-locality. The most well known such criteria take the form of violation of so-called ``Bell inequalities.'' Here we consider the simplest such inequality, namely the CHSH inequality.

For a pure state the CHSH inequality belongs to the class of Bell inequalities and can serve as an operational (experimental) criterion to discriminate between a product (local) and an entangled (non-local) state. Through a series of local measurements on many copies of their shared two qubits, Alice and Bob determine the expected value of 
\begin{align}
\mathcal{B}_{\rm CHSH}= {}&\vec{a}\cdot\vec{\sigma}\otimes \vec{b}\cdot\vec{\sigma}
+ \vec{a}^\prime\cdot\vec{\sigma}\otimes \vec{b}\cdot\vec{\sigma} 
\nonumber \\ & 
+ \vec{a}\cdot\vec{\sigma}\otimes \vec{b}^\prime\cdot\vec{\sigma} - \vec{a}^\prime\cdot\vec{\sigma}\otimes \vec{b}^\prime\cdot\vec{\sigma}
\end{align}
where $\vec{a}, \vec{a}^\prime, \vec{b}, \vec{b}^\prime$ are unit vectors in $\mathbb{R}^3$. Let 
\begin{equation}
    2\sqrt{M_{\rm AB}} \equiv \max_{\Vert\vec{a}\Vert=\Vert\vec{a}^\prime\Vert=\Vert\vec{b}\Vert=\Vert \vec{b}^\prime\Vert =1}{\rm Tr} \left[ \vert\Psi_{\rm AB}\rangle \langle \Psi_{\rm AB}\vert\mathcal{B}_{\rm CHSH} \right] 
    \label{eq:MAB}
\end{equation}
it is well known that $\vert\Psi_{\rm AB}\rangle$ is a product state if $M_{\rm AB} \leq 1$.
On the other hand if the pure state is entangled then the latter inequality is ``violated,'' the Bell states giving $M_{\rm AB}$ the maximum value $\sqrt{2}$. In view of this, a natural definition of an entanglement measure for pure states is
\begin{align}\label{eq:LAB}
    L_{\rm AB} = {\rm max}\Big(0, \frac{2 \sqrt{M_{\rm AB}} - 2}{2\sqrt{2} -2}\Big)
\end{align}

A generalization of the measure $L_{\rm AB}$ to general mixed two-qubit sates \eqref{eq:rho_gen} has been proposed in \cite{HORODECKI1995}. Consider the quantity $M(\rho)$ defined from \eqref{eq:MAB} but where $\vert\Psi_{\rm AB}\rangle\langle\Psi_{AB}\vert$ is replaced by a density matrix $\rho$. Define the CHSH-non-locality $L(\rho)$ as the quantity~\eqref{eq:LAB} where $M_{\rm AB}$ is accordingly replaced by $M(\rho)$. Remarkably CHSH-non-locality $L(\rho)$ can be computed explicitly and displays the following essential properties: 
\begin{itemize}
\item
We have $M(\rho)=\tau_1 +\tau_2$ the sum of the two largest eigenvalues (among three) of $T^\dagger T$ where $T = (t_{ij})$.
\item CHSH-local states naturally satisfy $M(\rho) \leq 1$. 
\item The maximum possible value of $M(\rho)=2$ is attained for pure Bell states.
\item For BDS from eq. \eqref{eq:rho_tuple} $T^\dagger T = {\rm diag}(t_1^2, t_2^2, t_3^2)$ so $\tau_1+\tau_2 = \Vert{\vec t}\Vert^2 - t_{\rm min}^2$ and \eqref{eq:LAB} becomes 
\begin{equation}
    L(\rho) =  \max\Big(0, \frac{\sqrt{\norm{\vec{t}}^2-t^2_{\rm min}} -1}{\sqrt{2} -1}\Big),
    \label{eq:non_locality}
\end{equation} 
where $t_{\rm min}= \min(|t_1|, |t_2|, |t_3|)$
\end{itemize}
This last formula is our main interest here. 
Non locality vanishes in the region $\{\vec{t} \mid \Vert \vec{t}\Vert^2 - t_{\rm min}^2 \leq 1\}$ which is just {\em the convex region corresponding to the intersection of three unit cylinders oriented along the main axes}. This region obviously contains the unit ball $\Vert\vec{t}\Vert \leq 1$, which in turn contains also the octahedron $\mathcal{O}$. The  common points are the $6$ vertices of $\mathcal{O}$ on the coordinate axes. 

Therefore states displaying CHSH-non-locality are also non-separable (or entangled). But not all non-separable states display CHSH-non-locality.

\begin{figure}[h]
		\centering
	    \includegraphics[width=0.3\textwidth, trim = {2cm 3cm 2cm 3cm}]{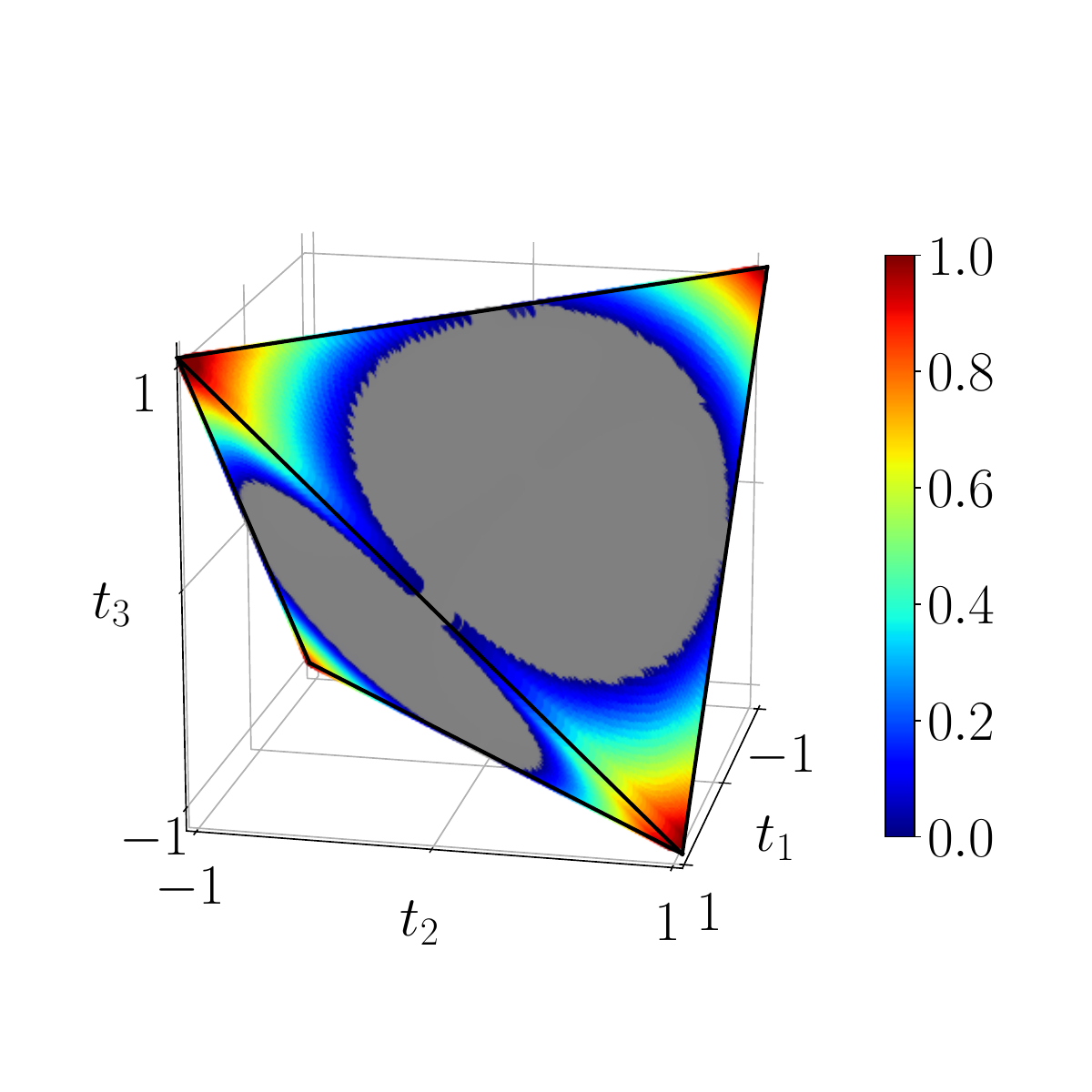}
		\caption{{\small CHSH-non-locality of BDS $L(\rho)$ given by \eqref{eq:non_locality}, calculated from the noiseless simulation of compact circuit (\cref{fig:G-compact}). In the gray region CHSH-non-locality identically vanishes.}}
		\label{fig:CHSH-non-locality}
\end{figure}

\Cref{fig:CHSH-non-locality} displays the measure $L(\rho)$ of CHSH-non-locality in the tetrahedron, computed using the simulation circuit of \cref{fig:G-compact}. This corresponds exactly to the analytical result \eqref{eq:non_locality}. We see that CHSH-non-locality vanishes in an ``inflated'' unit ball (intersection of three unit cylinders) which contains the octahedron $\mathcal{O}$, and is non-zero close to the four corners of the tetrahedron $\mathcal{T}$. At the extremal points corresponding to pure Bell states it reaches its maximum as expected.

\subsubsection{Steering} 

The notion of steering goes back to one of the most paradoxical aspects of quantum mechanics discussed by EPR and Schr\"odinger, but was formulated only recently \cite{Wiseman-et-al2007,Jones-et-al2007}. It is the ability that A has, by making only local measurements, to prepare or ``steer'' the state of party B. For example, for a pure Bell state $\vert \beta_{00}\rangle$ if A measures its qubit in the basis $\{\vert+\rangle, \vert -\rangle\}$ and obtains $\vert+\rangle$, then B's qubit is ``steered'' 
to $\vert +\rangle$. Of course this does not imply signaling and B is completely oblivious to the actions of A, his description of his qubit by his reduced density matrix remaining valid, unless he receives information sent by A. In order to convince B that his state has been steered by A, B must receive information from A and then do appropriate tests by local measurements on his side. 

The notion of steering for mixed states was formalized precisely in \cite{Wiseman-et-al2007} and seen as an intermediate between non-separability and Bell-non-locality. Roughly speaking, we say that A has the ability to steer the state of B if, after having received A's information, B cannot explain the results by a local-hidden-state (LHS) model (A would not be able to steer a local-hidden-state on B's side). Again, it is quite difficult to test all possible measurement situations. For example, even when restricting to dichotomic measurements we could imagine that A steers B's state by using only two types of measurements (2-steering), or three types of measurements (3-steering), and so on. A general measure for the steerability of two qubit states has been found \cite{Costa}. Just as for Bell-non-locality, sufficient criteria have been derived for assessing steerability of a state, and in general they take the form of inequalities \cite{PhysRevA.80.032112,Cavalcanti:15,steering_review}, but for two-qubit states one has a steering measure \cite{Costa}. 

Steerability of BDS has been discussed before \cite{quan2016steering,KuNoriPRA2018}. It turns out that BDS are ``2-steerable'' if and only if they are CHSH-non-local \cite{quan2016steering,Costa,PhysRevA.94.032317} (in fact \cite{PhysRevA.94.032317} shows this is true for all $2$-qubit mixed states). They are ``3-steerable'' as long as $\Vert\vec{t}\Vert \geq 1$ \cite{quan2016steering}, \cite{Costa}. Thus one can consider the measure \cite{Costa} of 3-steerability which distinguishes 3-steerability from CHSH-nonlocality, namely,
\begin{equation}
    S_3(\rho) = \max\Big(0, \frac{\norm{\vec{t}}-1}{\sqrt{3} -1}\Big),
    \label{eq:steering}
\end{equation}
where $\vec{t} = (t_1, t_2, t_3)$ is defined by \eqref{eq:param_prob}, and the factor $\sqrt{3}$ comes from the maximum violation of steering inequality ($\sqrt{n}$ for $n=2,3$ measurements per site \cite{quan2016steering}). 
We see that 3-steering $S_3(\rho)$ vanishes in the intersection of the sphere of radius one $\Vert\vec{t}\Vert \leq 1$ with $\mathcal{T}$. This sphere contains $\mathcal{O}$. Therefore non-zero 3-steering implies non-zero negativity and non-separability. But the reciprocal is not necessarily true.

\begin{figure}[h]
		\centering
		\includegraphics[width=0.3\textwidth, trim = {2cm 3cm 2cm 3cm}]{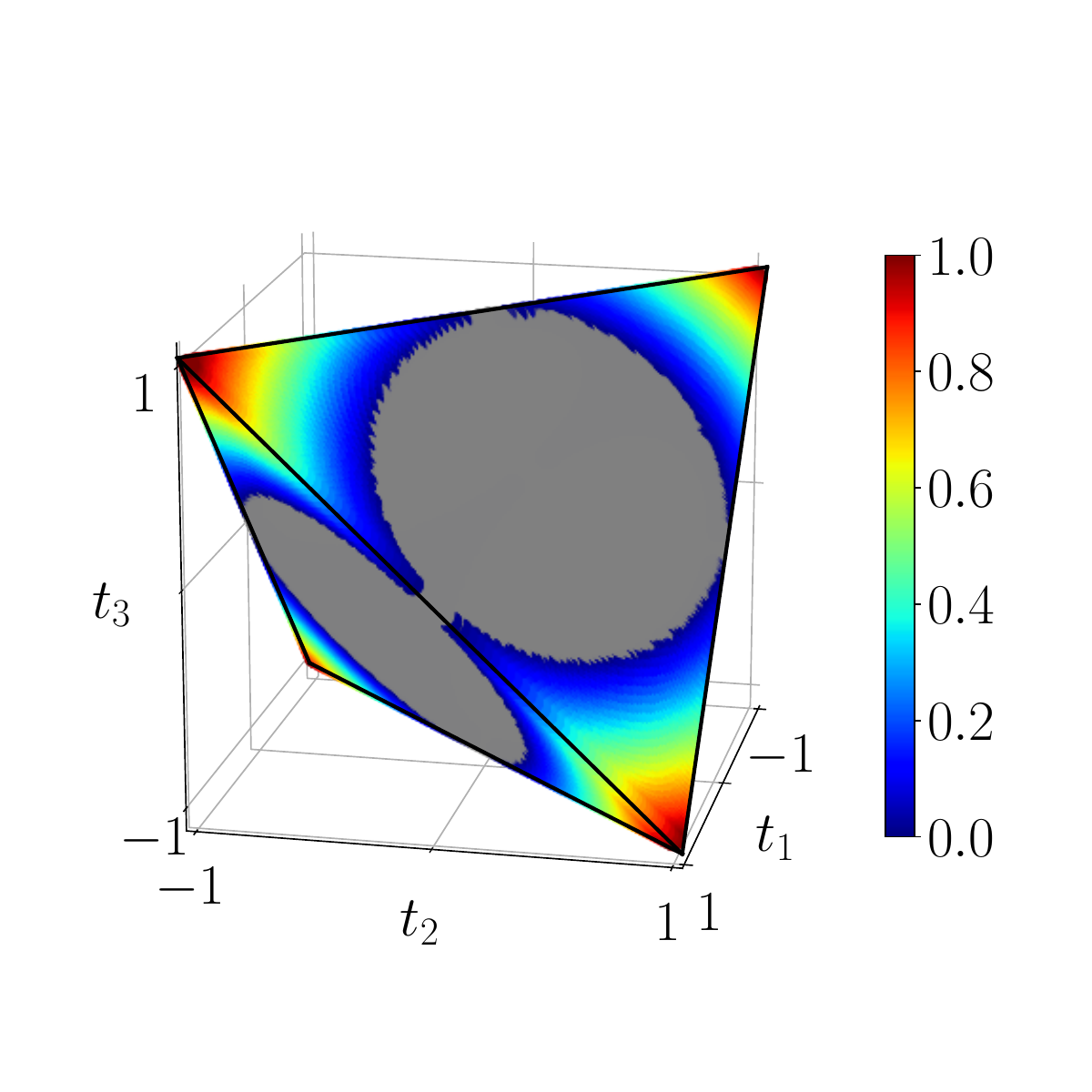}
		\caption{{\small $S_3(\rho)$, 3-steering of BDS, calculated from the noiseless simulation of compact circuit (\cref{fig:G-compact}). In the gray region 3-steering identically vanishes.}}
		\label{fig:3-Steering}
\end{figure}

\Cref{fig:3-Steering} displays the measure of 3-steering $S_3(\rho)$ in the tetrahedron, computed using the simulation circuit of \cref{fig:G-compact}. This agrees again with the analytical formula \eqref{eq:steering}. We see the intersection of the unit ball with the tetrahedron inside which steering vanishes.
States close to the four corners are 3-steerable and we observe that these domains are slightly bigger than the CHSH-non-locality ones. Unsurprisingly steering is maximized at the Bell states. 

\subsubsection{Hierarchy between quantum correlation measures: entanglement, steering and CHSH-non-locality}

The general question of hierarchy between different types of quantum correlations has been elusive due to the difficulty of defining good measures (see e.g. discussion of ordering in Refs.  \cite{Mirano2015} and  \cite{Mirano2004}). 

However there is a genuine hierarchy between non-separability, steering and non-locality which was first discussed for all projective measurements in the first seminal papers on steering \cite{Wiseman-et-al2007} (in terms of LHV and LHS models), using certain families of states among which Werner states. For generalized POVM measurements the corresponding proof has been given only recently \cite{Quintino2015}. The steering measure for two qubits proposed by \cite{Costa} and used above of course strictly obey this hierarchy. 

\begin{figure}[h]
	\centering
	\includegraphics[width=0.3\textwidth]{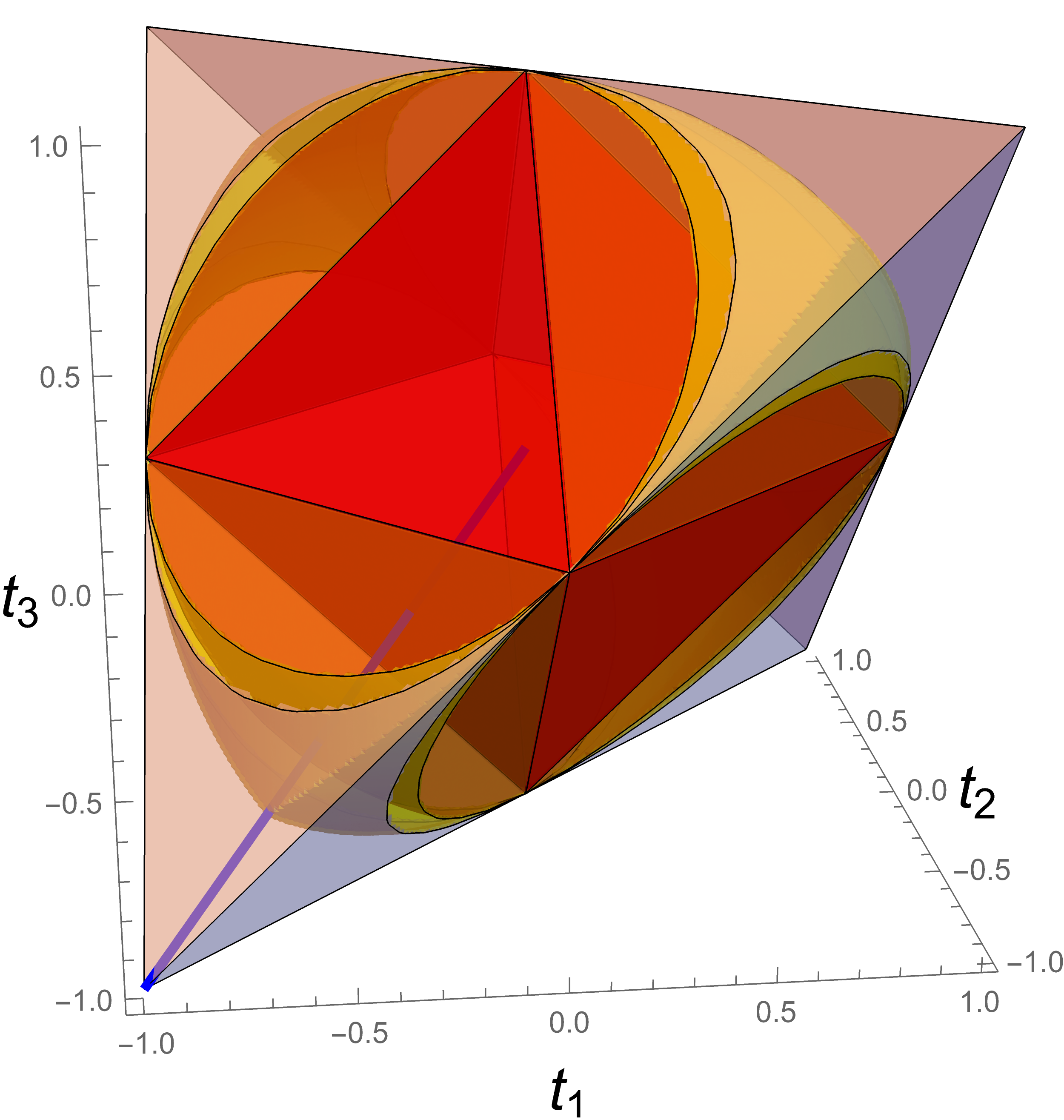}
	\caption{{\small Hierarchy of regions of separability (red), vanishing 3-steering
	(orange), vanishing CHSH-non-locality (yellow) and the rest of the BDS tetrahedron.
	The Werner line is also shown.}}
	\label{fig:hierarchy}
\end{figure}

One can explicitly illustrate this here for 3-steering inside the full tetrahedron of BDS, as shown in \cref{fig:hierarchy}. From \eqref{eq:steering}, 3-steering $S_3(\rho)$ vanishes in the unit ball. This sphere is strictly bigger than the octahedron $\mathcal{O}$, so there exist non-separable entangled states that do not exhibit steering. Similarly from \eqref{eq:non_locality} CHSH-non-locality $L(\rho)$ vanishes in the region corresponding to the intersection of three unit cylinders (oriented along the main axes), region which contains the unit ball. Thus there exist states that exhibit 3-steering and are not CHSH-non-local (do not violate the CHSH inequality). 

Summarizing, the BDS nicely exhibit the following hierarchy: i) states violating the CHSH inequality exhibit 3-steering; ii) states exhibiting steering are non-separable or entangled. The neighborhood of the four corners of the tetrahedron $\mathcal{T}$ display all three properties, and for the Bell states all these entanglement measures are maximal.

Finally we recall that, as pointed out above, the sets of 2-steerable and CHSH-non-local BDS are identical.

\subsection{Discord for BDS}\label{sec:discord}
It is not obvious how to quantify non-classicality of quantum correlations, which are distinct from entanglement.
Ollivier-Zurek \cite{Quantum_Discord_Ollivier_Zurek2001} approached this problem by introducing information  theoretical measures, and introduced ``quantum discord'' as the discrepancy between two quantum forms of mutual information. This notion has a few shortcomings, for example it applies only to bipartite systems treated asymmetrically, and other measures of non-classicality have been proposed since then. Among them, one of the most natural and conceptually clear, is the ``relative entropy of discord'' \cite{Modi2010}. This notion is based on a distance measure between a general multipartite state and its closest ``classical state.'' In this paragraph we first shortly review these two notions of discord and refer the reader to the review \cite{Modi2012} for a more complete discussion of particular aspects of these and other related notions of quantum correlations. We adopt the terminology used in this review and investigate more in detail the relationship between quantum discord and asymmetric relative entropy of discord, for which we find a general inequality.

For BDS, as we will see in the next paragraph, we find that quantum discord and asymmetric relative entropy of discord become one and the same. This however, according to our inequality, is not even true for general two-qubit systems, and one can only assert that quantum discord is smaller or equal than asymmetric relative entropy of discord.

\subsubsection{Quantum discord}

We explain the information theoretical point of view of reference \cite{Quantum_Discord_Ollivier_Zurek2001}.
The quantum mutual information of a bipartite mixed state $\rho_{AB}$
is defined as
\begin{align}
    (\ln 2)\, \mathcal{I}(A;B)  & = S(\rho_A\otimes\rho_B) - S(\rho_{AB}) \nonumber \\ &
    = S(\rho_A) + S(\rho_B) - S(\rho_{AB})
    \label{eq:MI1}
\end{align}
This a measure of total correlation which is 
the closest analog to the fundamental expression of Shannon's mutual information $I(X;Y) = H(X)+H(Y) -H(X,Y)$ defined 
for two random variables $X$ and $Y$ \cite{Cover:2006:EIT:1146355}. But Shannon's mutual information can also be written as 
$I(X;Y) = H(X) - H(X\vert Y)$, i.e., the difference between Shannon's entropy of $X$ and the conditional entropy of $X$ when $Y$ is observed \cite{Cover:2006:EIT:1146355}. We seek a quantum analog of this second form of mutual information.
Imagine that party $B$ makes local measurements with a complete set of orthonormal projectors $\{\mathcal{I}\otimes \Pi_{B}^k\}$ without recording the measurement outcomes (here we restrict ourselves to projective measurements instead of the more general definition involving POVM). The post-measurement description of the global state is 
$
\rho_{AB}^{\{\Pi_{B}^k\}} = \sum_k (\mathcal{I}\otimes \Pi_{B}^k) \rho_{AB} (\mathcal{I}\otimes \Pi_{B}^k)
$,
the one of party $B$ is $\rho_{B}^{\{\Pi_{B}^k\}} = \sum_k \Pi_{B}^k \rho_{B} \Pi_{B}^k$,
and the one  of party $A$ remains equal to $\rho_A$ (which is compatible with no-signaling). In this situation the mutual information after the measurement is defined as 
\begin{align}
    (\ln 2)\, \mathcal{C}(A;B\vert & \{\Pi_{B}^k\})  = 
    S(\rho_A\otimes\rho_B^{\{\Pi_{B}^k\}}) - S(\rho_{AB}^{\{\Pi_{B}^k\}})
    \nonumber \\ &
    = S(\rho_A) + S(\rho_B^{\{\Pi_{B}^k\}}) - S(\rho_{AB}^{\{\Pi_{B}^k\}})
    \label{eq:MI2}
\end{align}
This quantity has been called the ``classical correlation.'' It bears two striking differences with its classical analog. Firstly it depends on the measurement basis (a non-issue in the classical case) and secondly it is not the same when A is measured instead of B (whereas in the classical case we have $H(X) - H(X|Y) = H(Y) - H(Y|X)$).

While in the classical case \eqref{eq:MI1} and \eqref{eq:MI2} both reduce to Shannon's $I(X; Y)$, they are not equal for quantum systems. 
The {\it quantum discord} is defined as the difference between \eqref{eq:MI1} and the maximum of \eqref{eq:MI2} over all possible measurement basis, i.e.
\begin{align}
\mathcal{D}(\rho_{AB})  = \mathcal{I}(A;B) - \mathcal{C}(A;B)
\label{eq:DefinitionDiscord} 
\end{align}
where $\mathcal{C}(A;B) = \max_{\{\Pi_{B}^k\}}\mathcal{C}(A;B\vert \{\Pi_{B}^k\})$. 

To summarize, $\mathcal{I}(A;B)$ is interpreted as the amount of total correlation between the two parties, $\mathcal{C}(A;B)$ as the amount of classical correlation, and $\mathcal{D}(\rho_{AB})$ as {\em the amount of non-classical correlation}. 

It is a theorem that all three quantities are non-negative but in general not much more can be said about the relative magnitude of classical and quantum correlations. Clearly $\mathcal{I}(A;B)$ is symmetric under exchange of $A$ and $B$, but this is not the case 
for $\mathcal{C}(A;B)$ and $\mathcal{D}(A;B)$ (one sometimes speaks of right-discord when B is measured and left-discord when A is measured).
But note that if the two parties are identical systems these quantities are symmetric. This is the case for BDS.

\subsubsection{Asymmetric relative entropy of discord}

We first explain the hierarchical point of view of reference \cite{Modi2010} which is based on relative entropy as a distance measure, and which presents discord as a distance to the closest classical state. Although we restrict here to bipartite systems the discussion readily extends to multipartite situations. Classical states are defined as statistical ensembles of perfectly distinguishable orthonormal product states 
$\vert k_A\rangle\otimes \vert k_B\rangle = \vert k_A k_B\rangle$, that is 
\begin{align}\label{eq:classical-state}
\chi = \sum_{k_A, k_B} p_{k_Ak_B} \vert k_A k_B\rangle \langle k_A k_B\vert
\end{align}
and $p_{k_A k_B}$ is a set of probabilities summing to one.
Let us call $\mathcal{C}$ the set of all possible classical states. The relative entropy of discord is defined as 
\begin{align}\label{eq:distance-discord}
    (\ln 2)\, D(\rho_{AB}) = \min_{\chi\in\mathcal{C}} S(\rho_{AB}|| \chi)
\end{align}
where the relative entropy is (by definition) 
$
S(\rho_{AB}|| \chi) = {\rm Tr} \rho_{AB} \ln \rho_{AB} - {\rm Tr} \rho_{AB} \ln \chi
$.
This quantity obviously treats A and B symmetrically, and as such it is not equivalent to \eqref{eq:DefinitionDiscord}. 

In \eqref{eq:DefinitionDiscord} the root of the asymmetry between $A$ and $B$ lies in the amount of classical correlation \eqref{eq:MI2}, which is measured only with respect to the $B$ system ($B$ plays here the role of $A$ in original papers \cite{Modi2010,henderson2001classical,Quantum_Discord_Ollivier_Zurek2001}). Therefore to establish a meaningful link between the two kinds of discord it is first necessary to minimize in both cases on the same asymmetric statistical ensemble, consisting of orthonormal product states with respect to $B$ only, whose elements read:
\begin{align}\label{eq:Bclassical-state}
\chi' = \sum_{k} (\mathcal{I} \otimes \vert k\rangle \langle k\vert)\, \rho\, (\mathcal{I} \otimes \vert k\rangle \langle k\vert) \in \mathcal{C'}
\end{align}
where {\em both} $\rho$ and the set $\{\vert k\rangle \langle k\vert \equiv \Pi_{B}^k\}$ are free parameters defining the ensemble $\mathcal{C'}$. We recognize $\chi'$ as possible post measurement states $\rho_{AB}^{\{\Pi_{B}^k\}}$. The corresponding relative entropy of discord then reads
\begin{align}\label{eq:Bdistance-discord}
    (\ln 2)\, D'(\rho_{AB}) = \min_{\chi'\in\mathcal{C'}} S(\rho_{AB}|| \chi')
\end{align}
whilst standard discord \eqref{eq:DefinitionDiscord} reads
\begin{align}
(\ln 2)\, \mathcal{D}(\rho_{AB})  = S(\rho_B) - S(\rho_{AB}) + \min_{\{\Pi_{B}^k\}} S(\rho_A|\Pi_{B}^k)
\label{eq:BDefinitionDiscord} 
\end{align}
where $S(\rho_A|\Pi_{B}^k)$ is the conditional entropy expressed as $\sum_{k_B} p_{k} S(\Pi_{B}^k \rho_{AB} \Pi_{B}^k / p_{k})$ with $p_{k} = \Tr(\Pi_{B}^k \rho_{AB} \Pi_{B}^k)$.

Ref. \cite{Modi2010} shows that the two forms of discord are related when one does {\it not} minimize with respect to the measurement basis.
Indeed for any fixed set $\{\vert k\rangle\}$ we define
$\mathcal{D}_{\{\vert k\rangle\}}(\rho_{AB})$ and $D^\prime_{\{\vert k\rangle\}}(\rho_{AB})$ as the quantities in \eqref{eq:Bdistance-discord} and 
\eqref{eq:BDefinitionDiscord} {\it without} the minimizations over $\{\vert k\rangle\}$. Note that to obtain $D^\prime_{\{\vert k\rangle\}}(\rho_{AB})$ one still has to minimize over the $\rho$ dependence of $\chi^\prime$. 
Then the following remarkably simple relation holds,
\begin{align}\label{eq:L-modi-et-al}
(\ln 2)\, [\mathcal{D}_{\{\vert k\rangle\}} - D'_{\{\vert k\rangle\}}](\rho_{AB}) = S(\rho_A\otimes\rho_B) - S(\pi_{\chi'_{\rho_{AB}}}) ,
\end{align}
where $\pi_{\chi'_{\rho_{AB}}}$ is the product of the two reduced density matrices associated to $\chi'_{\rho_{AB}}$, which itself is defined as the asymmetric classical state in $\mathcal{C}'$ which minimizes $D(\rho_{AB}|| \chi^\prime)$ when the minimization is carried on over $\rho$ only (for brevity's sake the $\{\vert k\rangle\}$-dependence of $\chi'_{\rho_{AB}}$ and $\pi_{\chi'_{\rho_{AB}}}$ is left implicit). 
Now, there is a modified version of theorem 2 of \cite{Modi2010} (with a similar proof) which states that, for any fixed orthonormal basis $\{\vert k\rangle\}$, the minimizer of $S(\rho_{AB}|| \chi')$ over $\rho$ is attained at $\rho=\rho_{AB}$, or equivalently at $\chi'_{\rho_{AB}} = \sum_{k} (\mathcal{I} \otimes \vert k\rangle \langle k\vert)\, \rho_{AB}\, (\mathcal{I} \otimes \vert k\rangle \langle k\vert)$. We note that this minimizer is such that $\Tr_A(\chi'_{\rho_{AB}})$ has eigenvectors $\vert k\rangle$ with eigenvalues $\langle k\vert \rho_B \vert k\rangle$. Moreover 
$\Tr_B (\chi'_{\rho_{AB}}) = \rho_A$, and we find
\begin{align}\label{eq:pichiab}
\pi_{\chi'_{\rho_{AB}}} =  \rho_A\otimes \sum_{k} \vert k\rangle  \langle k \vert \rho_B \vert k \rangle \langle k\vert .
\end{align}
It should be stressed that this expression, which will be useful later on, is {\it not} in general equal to $\rho_A\otimes \rho_B$ because $\vert k\rangle$ are eigenvectors of $\Tr_A(\chi'_{\rho_{AB}})$ only, moreover it still depends on the basis $\{\vert k\rangle\}$. The latter remark also holds for relationship \eqref{eq:L-modi-et-al}, which, as remarkable as it is, remains insufficient to directly relate the ``true'' discords $\mathcal{D}(\rho_{AB})$ and $D'(\rho_{AB})$ since they are defined by {\em independent} minimizations over $\{\vert k\rangle\}$.

We would now like to show that still it is possible to find a weaker relation between the two kinds of asymmetric discords in the form of a general and useful inequality. Consider the difference on the r.h.s of \eqref{eq:L-modi-et-al} which equals $S(\rho_B) - S(\sum_k \vert k\rangle \langle k\vert \rho_B\vert k\rangle \langle k \vert)$. We claim that this difference of entropies is {\it non-positive}. Thus \eqref{eq:L-modi-et-al} also is non-positive for all $\{\vert k\rangle\}$, which implies that $\mathcal{D}_{\{\vert k\rangle\}}(\rho_{AB}) \leq D'_{\{\vert k\rangle\}}(\rho_{AB}), \forall \, \{\vert k\rangle\}$, hence also
\begin{align}\label{eq:ineqAsDiscords}
\mathcal{D}(\rho_{AB}) \leq  D'(\rho_{AB}) \;.
\end{align}
This means that in general original discord cannot be larger than the corresponding asymmetric relative entropy of discord. To show the claim note that from the spectral decomposition $\rho_B = \sum_\beta \lambda_\beta \vert \beta \rangle \langle \beta \vert$ we have $S(\rho_B) = (\ln 2)\, H(\{\lambda_\beta\})$ and 
$S(\sum_k \vert k\rangle \langle k\vert \rho_B\vert k\rangle \langle k \vert) = (\ln 2)\, H(\{\sum_\beta |\langle k\vert \beta\rangle|^2 \lambda_\beta\})$ where $H$ is the classical Shannon entropy. It is a standard property of Shannon's entropy that it may only increase when a so-called doubly stochastic matrix (here $\vert \langle k\vert \beta\rangle|^2$) is applied to a probability distribution (see for example \cite{ash1990information} chap. 1, p. 26).

\subsubsection{Application to BDS}

Concerning BDS another fundamental result emerges from the previous discussion. 
Consider the r.h.s. of equation \eqref{eq:L-modi-et-al}: on one hand we already know that for BDS $\rho_A = \rho_B = \frac{1}{2}\mathcal{I}$, so $\rho_A \otimes \rho_B = \frac{1}{4}\mathcal{I} \otimes \mathcal{I}$, on the other hand equation \eqref{eq:pichiab} implies that $\pi_{\chi_{\rho_{AB}}^\prime}$ is also equal to $\frac{1}{4}\mathcal{I} \otimes \mathcal{I}$ for all $\{\vert k\rangle\}$. So one immediately sees that the r.h.s of equation \eqref{eq:L-modi-et-al} vanishes for all $\{\vert k\rangle\}$, and thus $\mathcal{D}(\rho_{AB})=D'(\rho_{AB})$ in the whole BDS tetrahedron. Therefore we conclude that {\em for all BDS} quantum discord and asymmetric relative entropy of discord are {\it equal} quantities.

To derive a concrete expression for the BDS quantum discord one should solve the optimization problems \eqref{eq:BDefinitionDiscord} and/or \eqref{eq:distance-discord}. As we just proved above, both problems have the same solution. Luo~\cite{Luo2008} was the first to solve \eqref{eq:BDefinitionDiscord} and gave explicit formulas for the mutual information, classical correlation and discord of BDS. 
 
From the original definition of BDS one notes that the eigenvalues of the density matrix for a point $(t_1,t_2,t_3)$ of $\mathcal{T}$ are given by \eqref{eq:param_prob}, which allows us to immediately write down $S(\rho_{AB})$. On the other hand $S(\rho_A)=S(\rho_B)=\ln 2$ for every point of $\mathcal{T}$. Therefore 
\begin{align}
    \mathcal{I}_{\rm BDS} = \frac{1}{4}\Big[ & (1-t_{1}-t_{2}-t_{3})\log_{2}(1-t_{1}-t_{2}-t_{3}) 
    \nonumber \\ &
    +(1-t_{1}+t_{2}+t_{3})\log_{2}(1-t_{1}+t_{2}+t_{3})
    \nonumber \\ &
    + (1+t_{1}-t_{2}+t_{3})\log_{2}(1+t_{1}-t_{2}+t_{3}) 
    \nonumber \\ &
    +(1+t_{1}+t_{2}-t_{3})\log_{2}(1+t_{1}+t_{2}-t_{3})\Big].
    \label{eq:MIW}
\end{align}
From~\cite{Luo2008} we have the remarkably simple result in terms of $t = \max(\vert t_1\vert, \vert t_2\vert, \vert t_3\vert)$
\begin{align}\label{eq:CCW}
\mathcal{C}_{\rm BDS} = 1 - h_2\Big(\frac{1+t}{2}\Big)
\end{align}
where $h_2(x) = - x\log_2 x  - (1-x) \log_2(1-x)$ is the binary entropy function. The quantum discord $\mathcal{D}_{\rm BDS}$ is just the difference of the two expressions 
\eqref{eq:MIW} and \eqref{eq:CCW}. 

\begin{figure}[h]
		\centering
		\includegraphics[width=0.3\textwidth, trim = {2cm 3cm 2cm 3cm}]{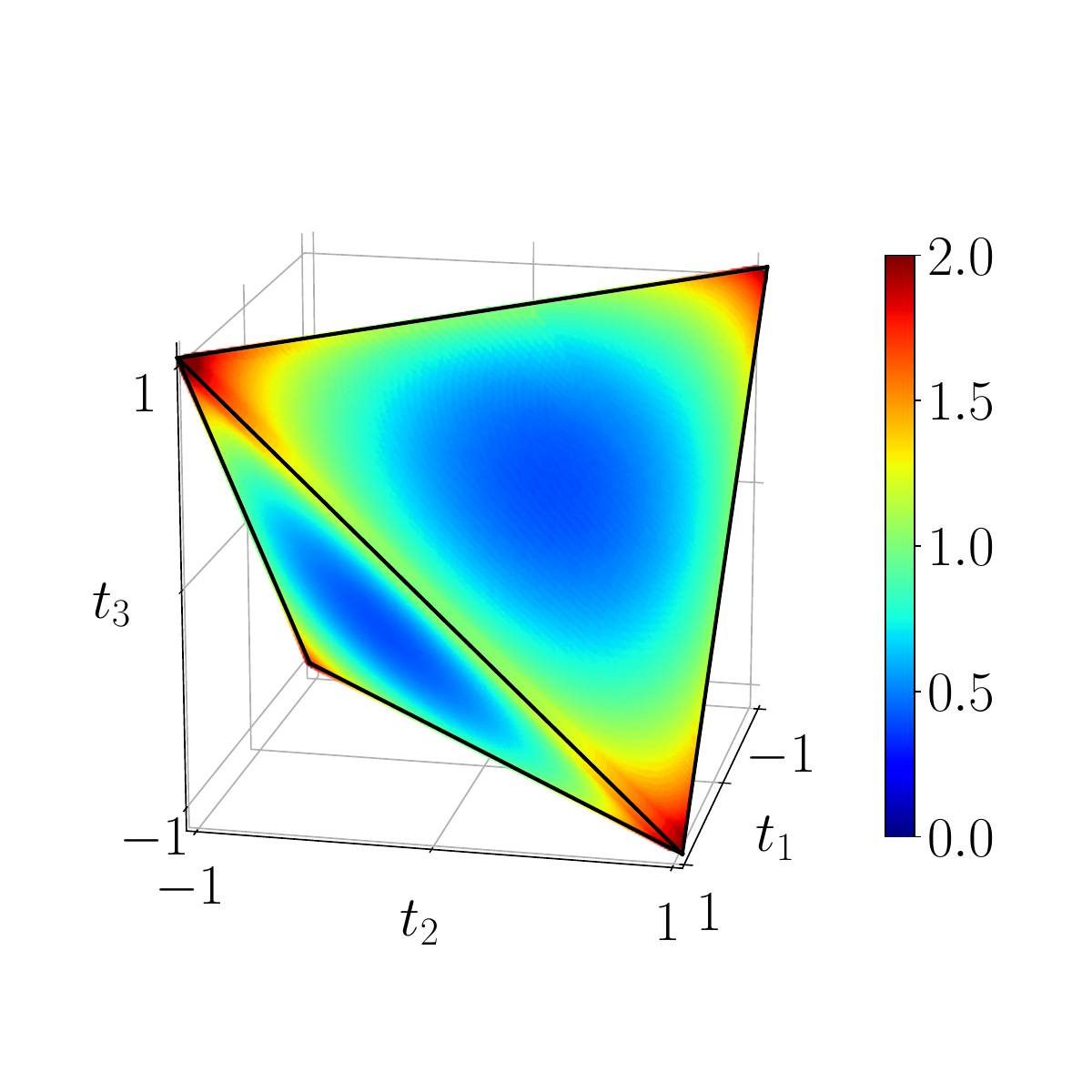}
		\caption{{\small Quantum mutual information $\mathcal{I}_{\rm BDS}$, calculated with \eqref{eq:MIW} from the noiseless simulation of compact circuit (\cref{fig:G-compact}). The range of mutual information is $[0,2]$.}}
		\label{fig:qmi_2}
\end{figure}

\begin{figure}[h]
		\centering
		\includegraphics[width=0.3\textwidth, trim = {2cm 3cm 2cm 3cm}]{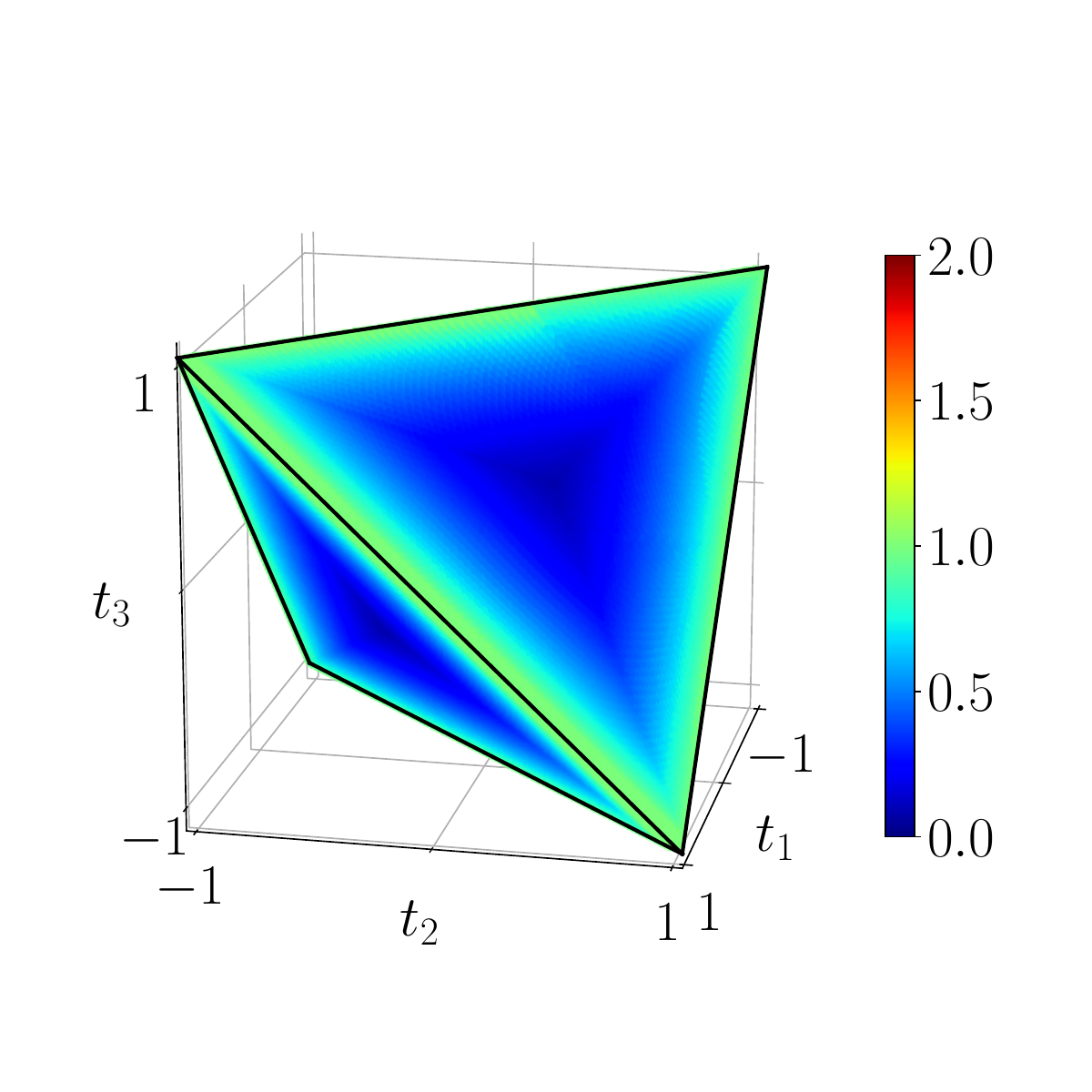}
		\caption{{\small Classical correlations $\mathcal{C}_{\rm BDS}$, calculated with \eqref{eq:CCW} from the noiseless simulation of compact circuit (\cref{fig:G-compact}). Plotted on the same range as mutual information.}}
		\label{fig:cc_2}
\end{figure}

\begin{figure}[h]
		\centering
		\includegraphics[width=0.3\textwidth, trim = {2cm 3cm 2cm 3cm}]{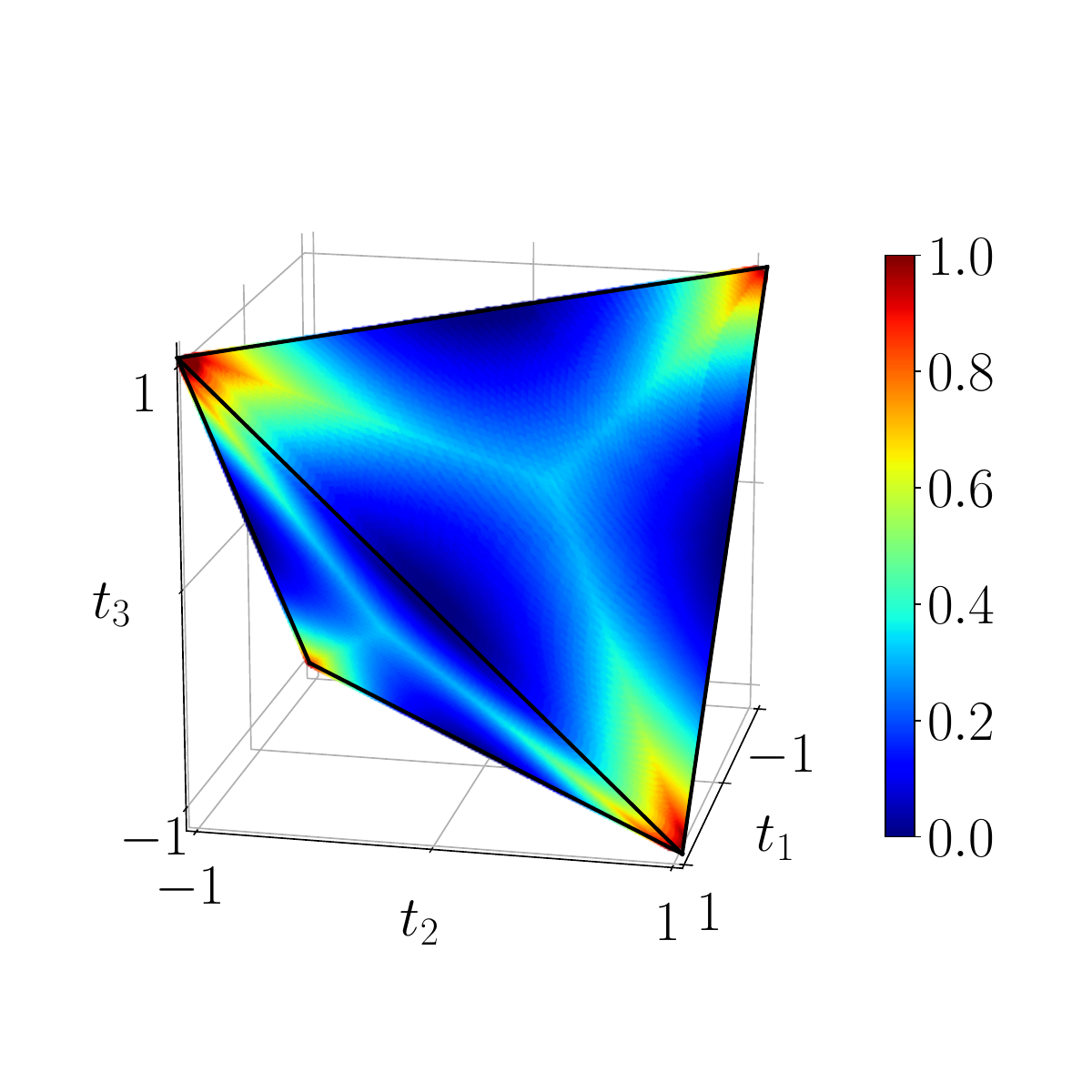}
		\caption{{\small Discord $\mathcal{D}_{\rm BDS} = \mathcal{I}_{\rm BDS} - \mathcal{C}_{\rm BDS}$, calculated with \eqref{eq:MIW} and \eqref{eq:CCW} from the noiseless simulation of compact circuit (\cref{fig:G-compact}). Here the range is the natural range $[0,1]$. Discord does not identically vanish on any extended domain, note the three-pointed star pattern on the faces.}}
		\label{fig:discord}
\end{figure}

Fig.~\ref{fig:qmi_2} shows the quantum mutual information of BDS computed on the tetrahedron with eq. \eqref{eq:MIW} , fig.~\ref{fig:cc_2} shows their classical correlation according to eq.\,\eqref{eq:CCW}, and fig.~\ref{fig:discord} displays their discord which is just their difference.

\subsection{The particular case of Werner states}\label{sec:wernermeasures}

\begin{figure}[h]
	\centering
	\includegraphics[width=0.4\textwidth, trim = {2cm 1cm 2cm 1cm}]{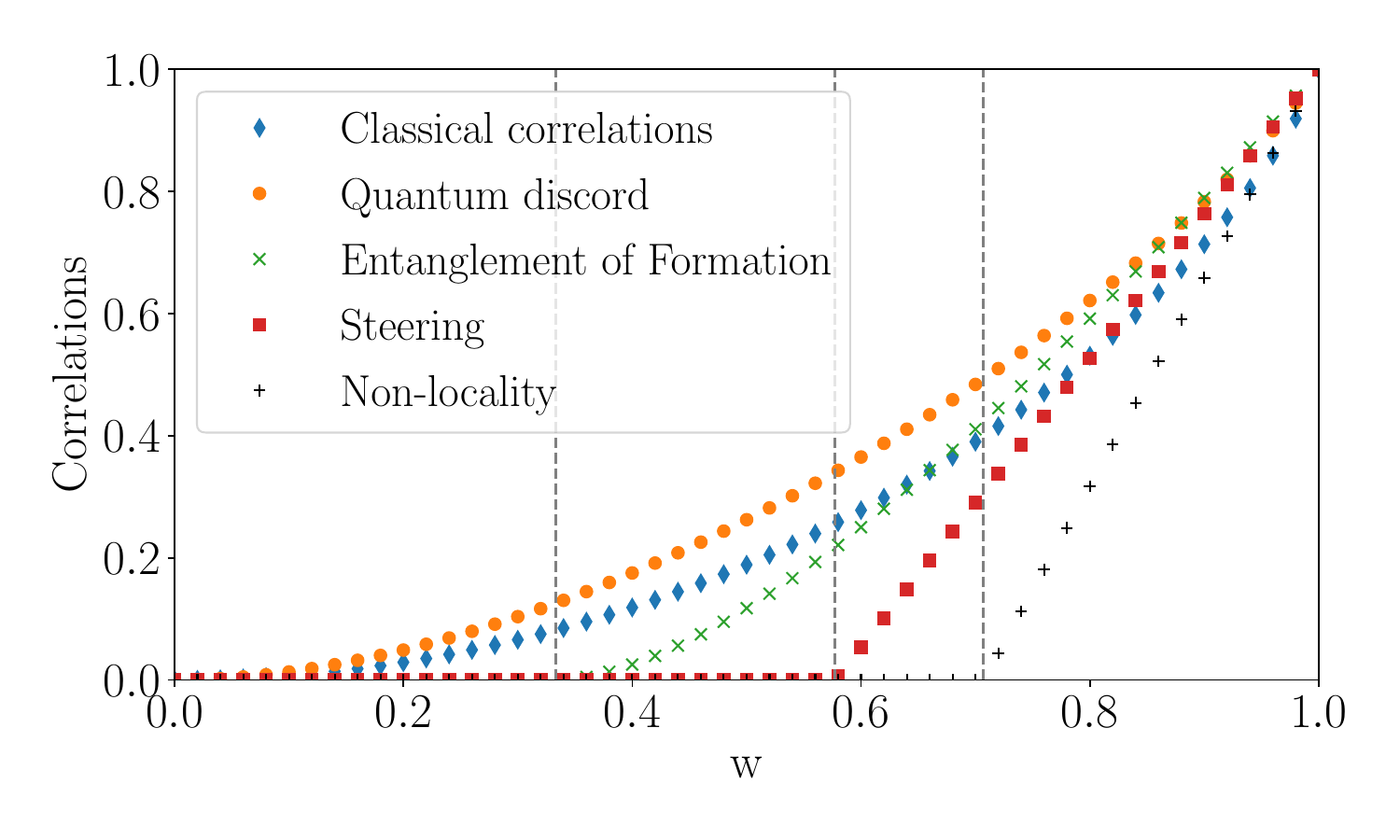}
	\caption{{\small Correlations of Werner states as a function of $w$ along the Werner line. The vertical bars mark the following critical values: entanglement of formation $E_{F}(\rho)$ vanishes for $w\leq \frac{1}{3}$ and states are separable, 3-steering $S_3(\rho)$ vanishes for $w\leq \frac{1}{\sqrt 3}$, CHSH-non-locality $L(\rho)$ and 2-steering both vanish for $w\leq \frac{1}{\sqrt 2}$. Discord $\mathcal{D}_W$ and classical correlations $\mathcal{C}_W$ are always positive, and discord is always bigger than classical correlations. Both are monotonously increasing on $[0,1]$.}}
	\label{fig:Wernerquantities}
\end{figure}

We recall that Werner states, defined in  \eqref{eq:rho_werner}, lie on the negative diagonal $t_1=t_2=t_3= -w$, $w\in [0, 1]$ of $\mathcal{T}$. Formulas for entanglement of formation $E_{F}(\rho)$, steering, CHSH-non-locality $L(\rho)$, 3-steering $S_3(\rho)$ and discord can be easily specialized on this line. The resulting quantities are plotted on \cref{fig:Wernerquantities}. We clearly observe the strict hierarchy discussed earlier: non-locality implies 3-steering which implies non-separability (or entanglement).

\begin{itemize}
    \item
{\it Non-separability and entanglement of formation $E_{F}(\rho)$.} 
The PPT criterion shows that Werner states are separable for $w\in [0, \frac{1}{3}]$ and display entanglement for $w\in (\frac{1}{3}, 1]$. The same threshold applies to concurrence (see \eqref{eq:conc})  and entanglement of formation (see \eqref{eq:wootters}. The details are given in appendix \ref{app:appendixwerner}.

\item 
{\it CHSH-non-locality.} CHSH-non-locality $L(\rho)$ vanishes for $w\in [0, \frac{1}{\sqrt 2}]$. Note that $\frac{1}{\sqrt 2}$ corresponds to the only points in the common intersection of the three unit cylinders oriented along the main axes.

\item
{\it Steering.} The threshold for 2-steering is identical to 
the one  of CHSH-non-locality \cite{quan2016steering,Costa,PhysRevA.94.032317}. On the other hand, 3-steering $S_3(\rho)$ vanishes for $w\in [0, \frac{1}{\sqrt 3}]$ and states with larger $w$ are 3-steerable. We point out that \cite{Wiseman-et-al2007} proved that Werner states cannot be replaced by a LHS model if an only if $w>\frac{1}{2}$ (this is the fundamental threshold below which Werner states are not steerable). 

\item
{\it Discord and classical correlation.}
From \eqref{eq:CCW}, the classical correlation is simply $\mathcal{C}_W=1-h_2(\frac{1-w}{2})$ and using \eqref{eq:MIW} we find the discord
\begin{align}
\mathcal{D}_W = {}&\frac{1}{4}(1-w)\log_{2}(1-w) -  \frac{1}{2}(1+w) \log_{2}(1+w)
\nonumber \\ &
+ \frac{1}{4}(1+3w)\log_{2}(1+3w)
\label{eq:discordWerner}
\end{align}
We note that discord is strictly bigger than classical correlation for all $w$ except $w=0$ and $1$.
\end{itemize}

\section{IBM Q results}\label{sect:IBMQ_results}

This section is divided in two parts. We first provide results obtained by simulations augmented with the noise model from IBM Q devices. Then we present experimental runs on real devices.

In this section we shall focus on only two types of quantities: first the fidelity of achievable density matrices with our circuits, this will allow an estimation of the error. Second we shall compute the corresponding classical correlations, quantum mutual information and discord. We have chosen these quantum correlations because they are the less trivial quantities, which do not vanish on any portion of the tetrahedron (except eventual singular points).

Fidelity will be displayed only in one-dimensional plots for Werner states on the Werner line, because the error for each value of the parameter $w$ can be easily visualized, as well as because the most important set of states in the tetrahedron are still covered. On the contrary classical correlations, quantum mutual information and discord will be displayed for the whole range of BDS in the tetrahedron.

The density matrices $\rho$ obtained from noisy simulations and/or experiments are reconstructed by Qskit tomography. Their accuracy can be measured thanks to the fidelity with respect to the theoretical density matrix $\rho^{\rm theo}$ of BDS
\begin{equation}
F(\rho,\rho^{\rm theo}) = \Big[ \Tr\big(\sqrt{\sqrt{\rho^{\rm theo}}\rho\sqrt{\rho^{\rm theo}}}\big)\Big]^2,
\label{eq:fid}
\end{equation}
The worst possible case would correspond to a maximally mixed reconstructed state $\rho\approx \frac{1}{4}\mathcal{I}\otimes \mathcal{I}$. On the Werner line this would yield
\begin{equation}
F_\mathrm{worst}(w) = \frac{1}{4}\big(\frac{3}{2}\sqrt{1 - w} +\frac{1}{2} \sqrt{1 + 3w}\big)^2.
\label{eq:ibmq_fid}
\end{equation}
This expression serves as a gross benchmark dotted line plotted on
\cref{fig:sim-circuit-comparison-noisy,fig:fidelity}. 

We first turn to simulations which give a first realistic expectation for experimental results, and which also allows to evaluate some circuits which cannot yet be realized on IBM Q. 

\subsection{Simulations with noise models from IBM Q quantum devices}
\label{sect:simulation_noise}

To compare the various quantum circuits that we have proposed in \cref{sect:circuits}, and specifically their expected performance in producing quantum correlations when subject to noise, we have run simulations using Qiskit \cite{Qiskit}. The source code for these is available in our GitLab repository \cite{gitlab-repo}. Each circuit is executed $2^{15}$ times, with noise simulated according to the Qiskit noise model \cite{NoiseModel}. Such noise models are generated semi-automatically by Qiskit based on the state of a real IBM Q device at the time of generation. As such, they are subject to change over time, and should in any case not be viewed as accurate representations of noise in real devices. They are, however, useful as a rough benchmark for comparing different quantum circuits.

\subsubsection{Fidelity in noisy simulations on the Werner line}

\begin{figure}[h]
   \centering
   \includegraphics[width=0.45\textwidth]{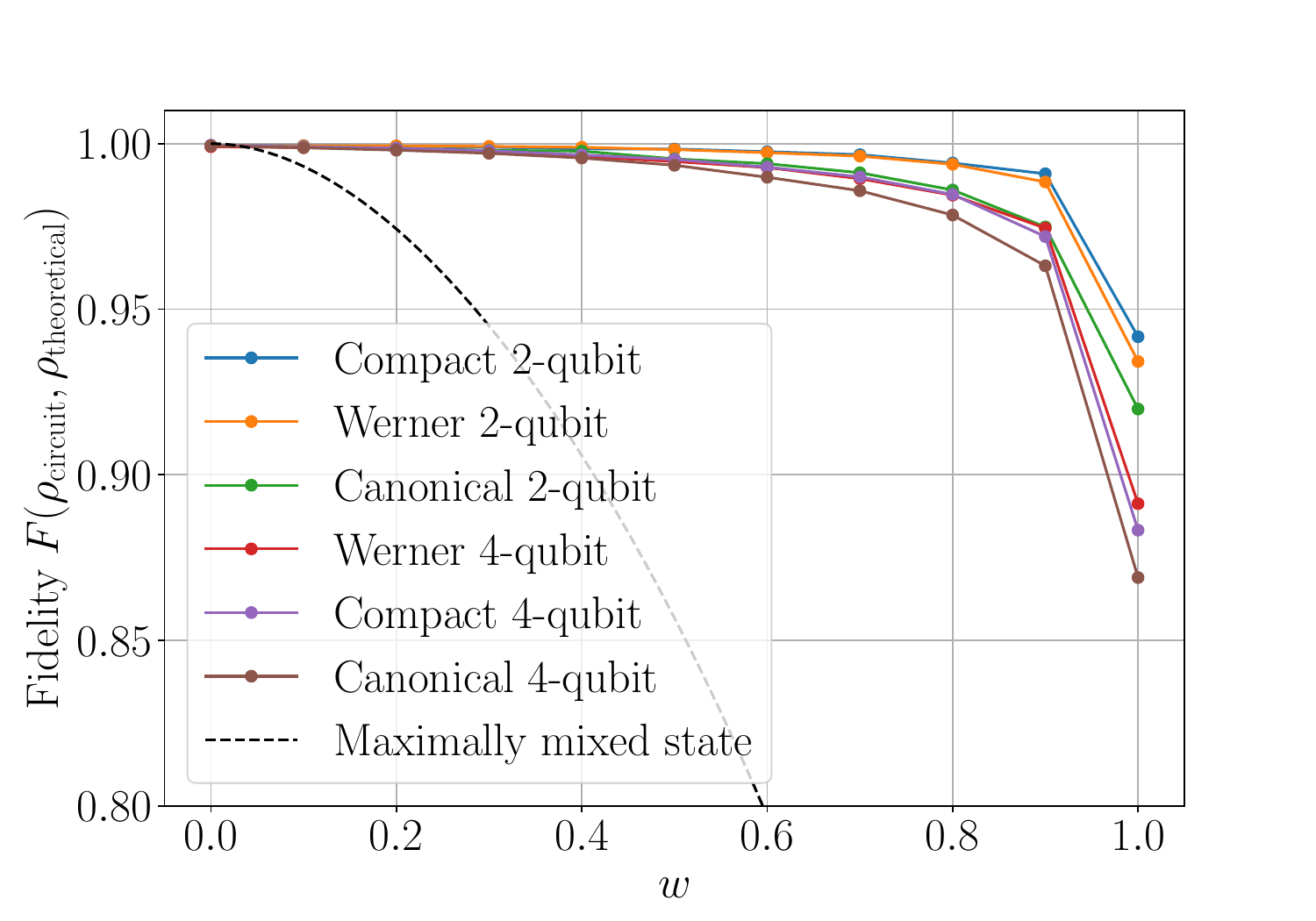}
    \caption{{\small Simulated fidelity curves $F(\rho^{\rm W\,circ},\rho^{\rm W\,theo})$ of Werner states as a function of the $w$ parameter, for density matrices produced by two-qubit and four-qubit versions of the circuits described in \cref{sect:circuits}. For comparison the black dashed line corresponds to a maximally mixed state. These results are based on tomography with $2^{15}$ shots under a Qiskit noise model generated for \texttt{ibmq\_athens} on 2021--05--16.}}
    \label{fig:sim-circuit-comparison-noisy}
\end{figure}

Fig.\,\ref{fig:sim-circuit-comparison-noisy} shows the fidelity $F(\rho^{\rm Wcirc},\rho^{\rm Wtheo})$ of Werner states $\rho^{\rm Wcirc}$ prepared with the different circuits described in \cref{sect:circuits}, with respect to the corresponding theoretical Werner density matrix $\rho^{\rm Wtheo}$, in presence of simulated Qiskit noise. The fidelity is computed from the density matrix of the output state, as empirically determined using our adaptation of Qiskit's built-in routine for quantum tomography; see \cref{sec:Qiskitimplementation}. As may have been expected, two-qubit circuits consistently outperform four-qubit circuits, and the ``compact'' circuit (\cref{fig:G-compact}) consistently outperforms the one based on canonical 3-sphere coordinates (\cref{fig:G-canonical}). Perhaps surprisingly, the two-qubit circuit specialized for Werner states (see \cref{fig:werner-circuit}) generally does not outperform the compact circuit.
	
\begin{figure}[h]
    \centering
    \includegraphics[width=0.3\textwidth, trim = {2cm, 3cm, 2cm, 3cm}]{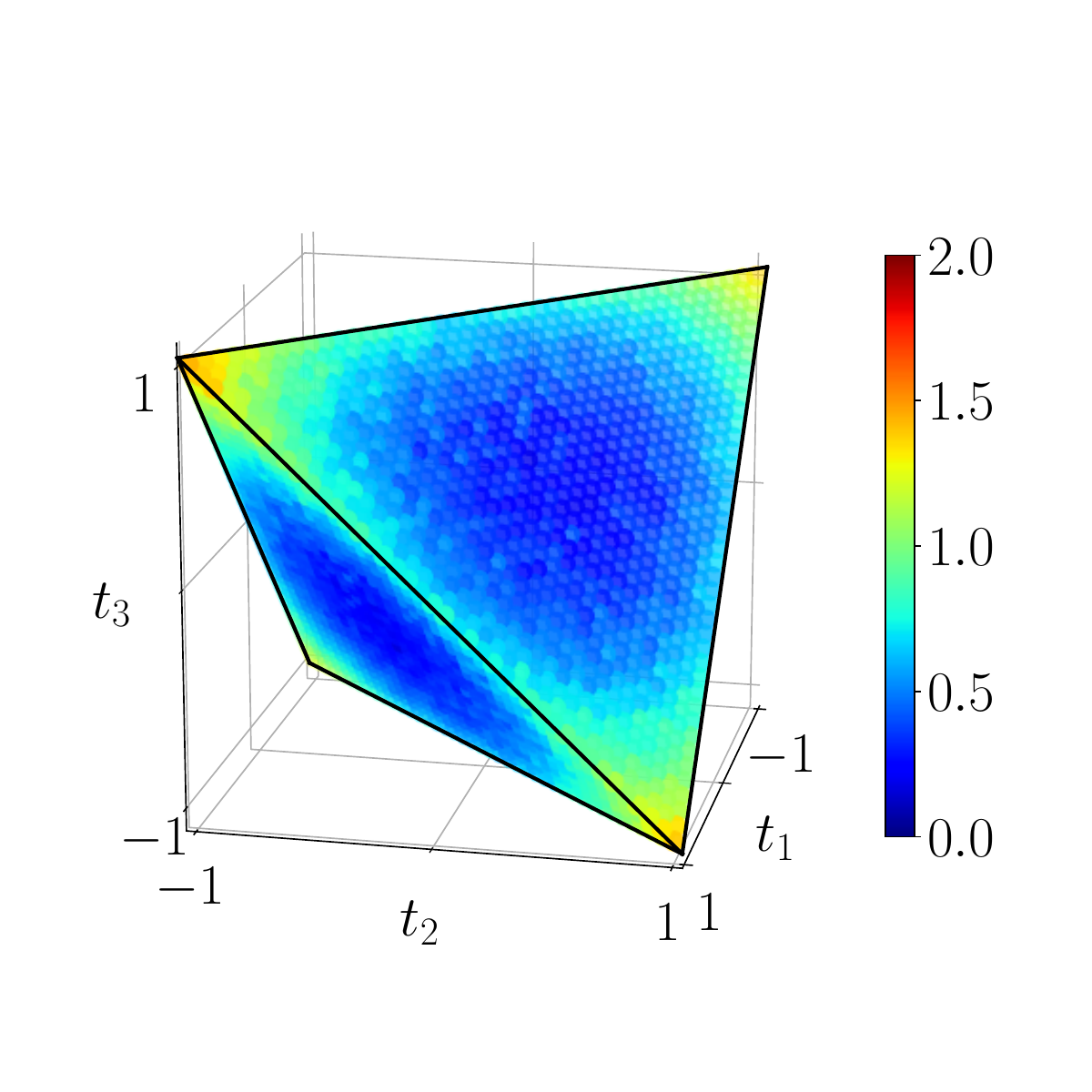}
    \includegraphics[width=0.3\textwidth, trim = {2cm, 3cm, 2cm, 3cm}]{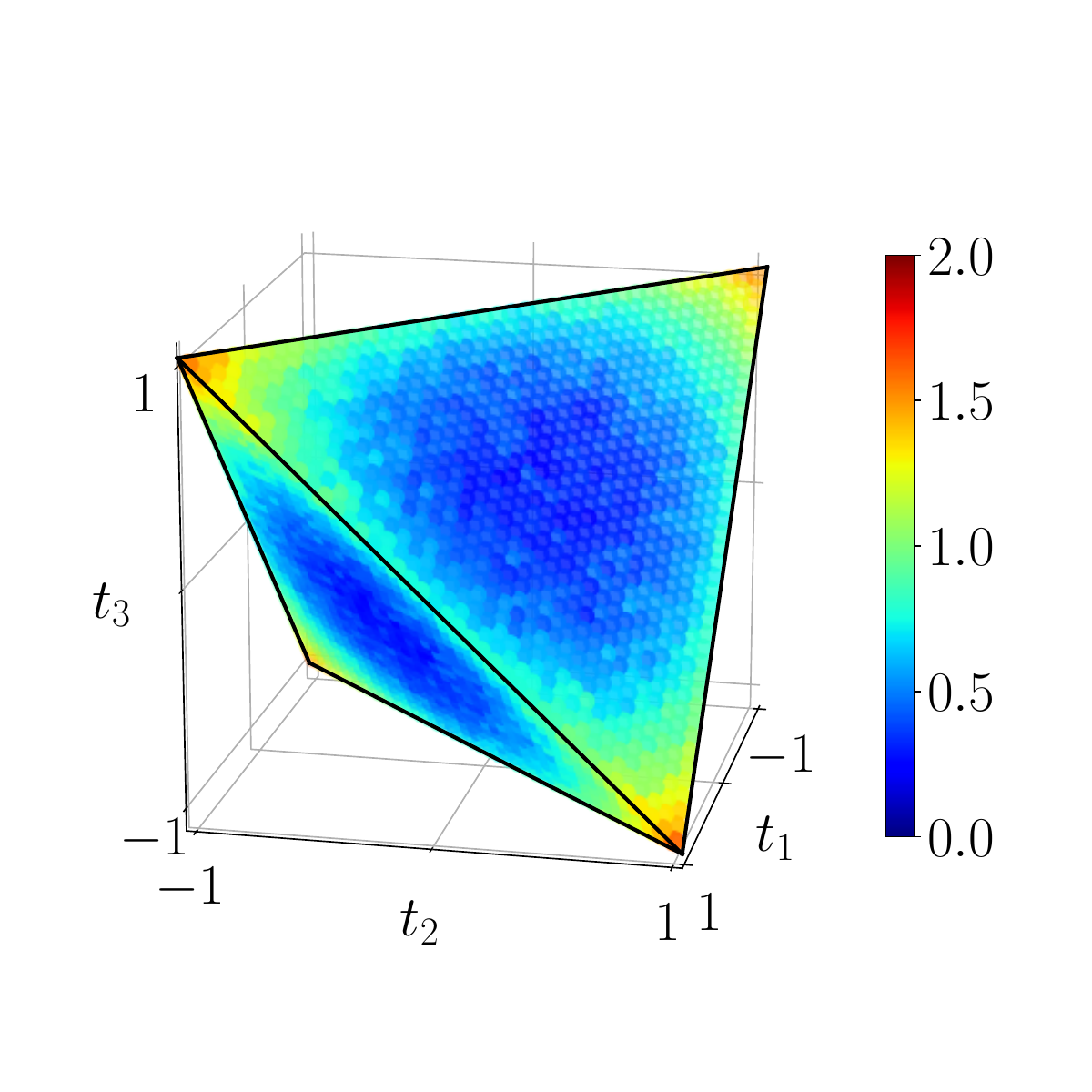}
    \caption{{\small Quantum mutual information $\mathcal{I}_{BDS}$ on its natural scale $[0, 2]$, as expected from noisy simulation of 4-qubit (top) and 2-qubit (bottom) circuits}.}
    \label{fig:noisy-qmi}
\end{figure}

\subsubsection{Noisy simulation in the whole tetrahedron: expected quantum mutual information and discord}

\begin{figure}[h]
    \centering
    \includegraphics[width=0.3\textwidth, trim = {2cm, 3cm, 2cm, 3cm}]{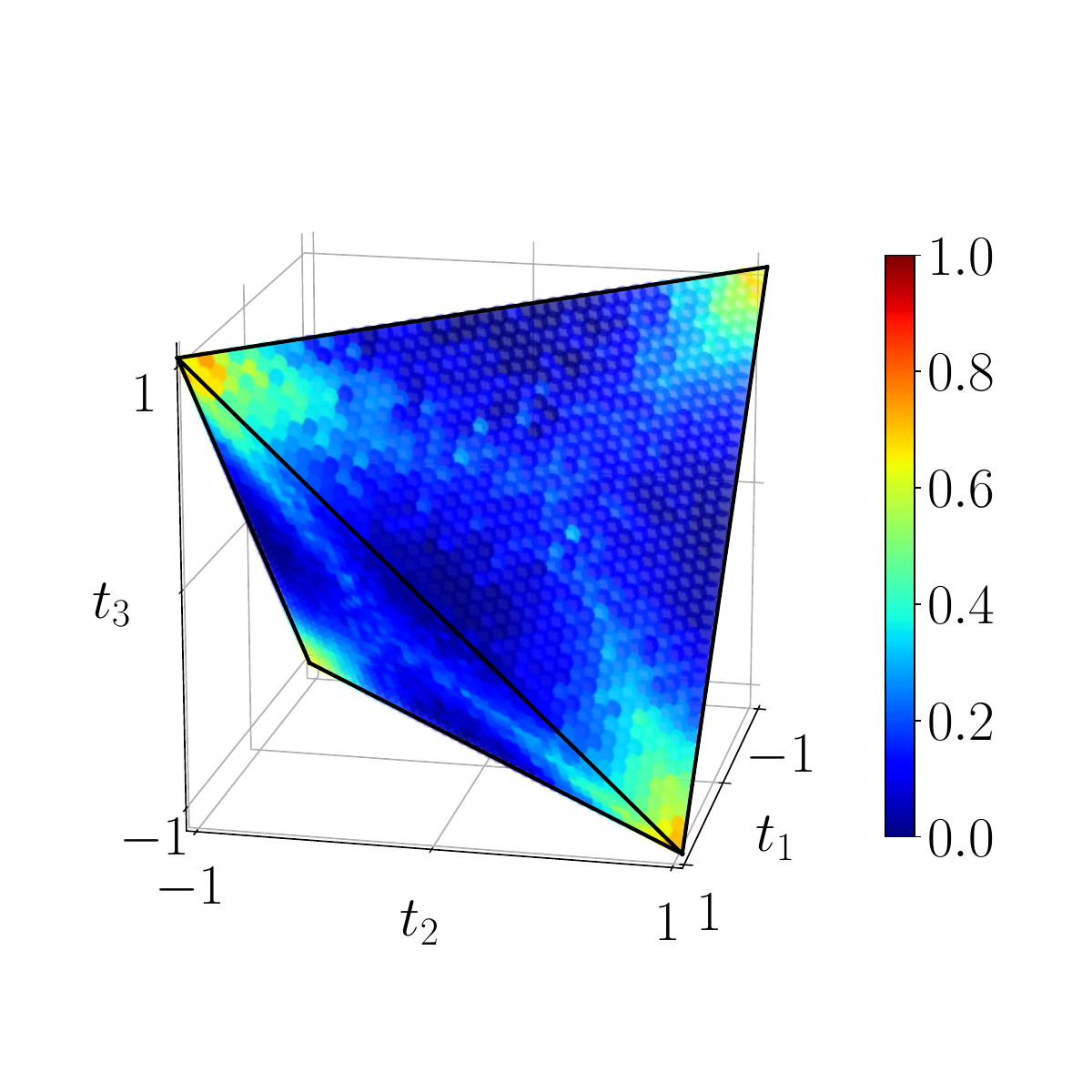}
    \includegraphics[width=0.3\textwidth, trim = {2cm, 3cm, 2cm, 3cm}]{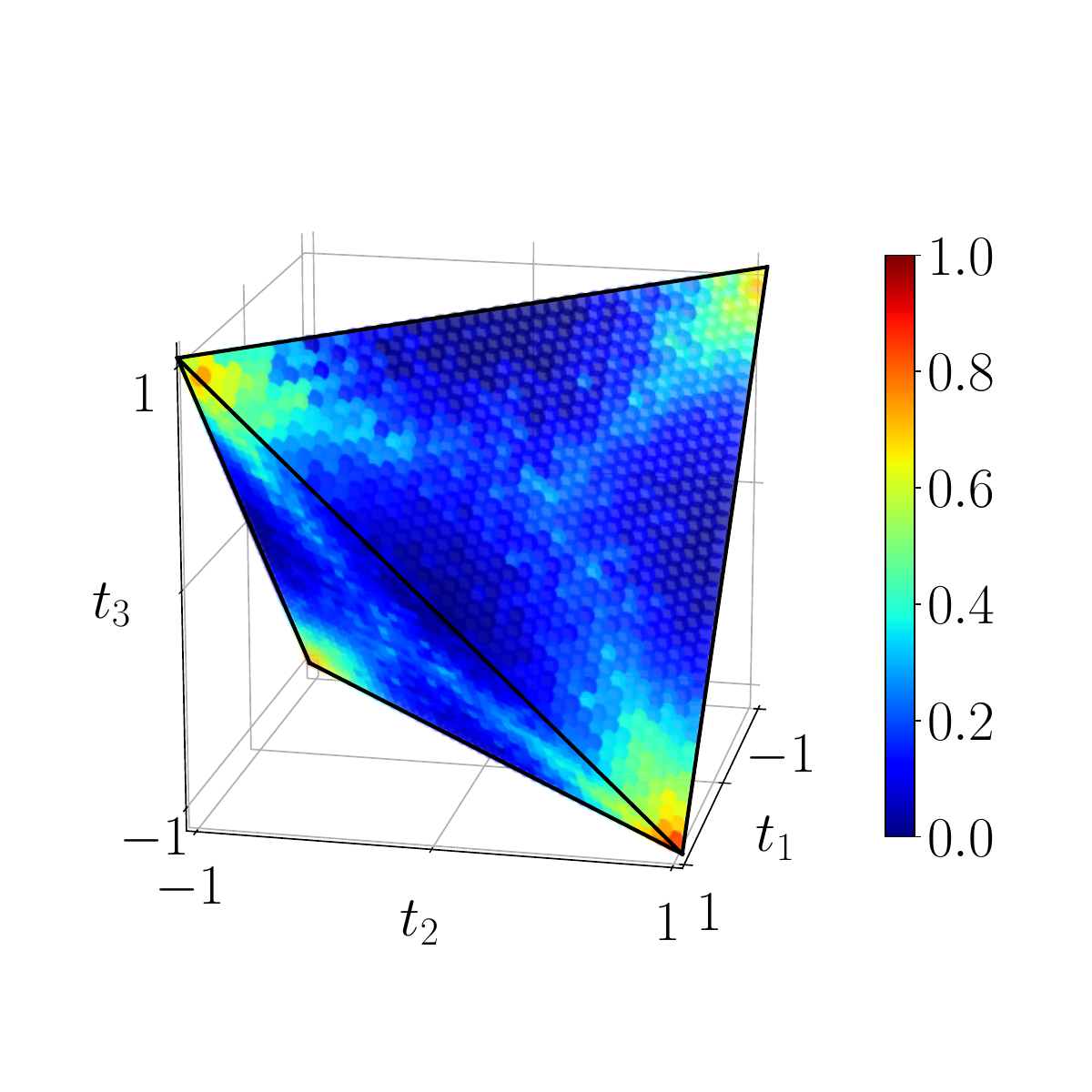}
    \caption{{\small Discord $\mathcal{D}_{BDS}$ on its natural scale $[0, 1]$, as expected from noisy simulation of 4-qubit (top) and 2-qubit (bottom) circuit. The three-pointed star pattern is visible, but deteriorating.}}
    \label{fig:noisy-discord}
\end{figure}

We have simulated the 4-qubit as well as 2-qubit circuits of \cref{sect:circuits} using a Qiskit noise model for the backend \texttt{ibmq\_london} (respectively generated on 2019--12--03 and 2019--12--10 with 1000 shots). Figs. \ref{fig:noisy-qmi} and \ref{fig:noisy-discord} display the result for the quantum mutual information and the discord on the whole tetrahedron. In general we observe that noise reduces these quantities to almost half their theoretical value close to the corners of the tetrahedron.
Interestingly, in the corners and along the edges we observe that the 2-qubit circuit is slightly more faithful to the ideal results of figs. \ref{fig:qmi_2} and~\ref{fig:discord} in \cref{sec:entanglementmeasures}. The same observations hold also for the classical correlation (not shown here). These results are consistent with the corresponding observations on the Werner line discussed in the previous paragraph.

\subsection{Experiments on IBM Q quantum devices}

We now turn to true quantum experiments on IBM Q. Some of the circuits which cannot yet be implemented are left out. The quantities measured and evaluated will be the same as in the simulations of the previous subsection, namely fidelity of the experimental density matrices (obtained by Qskit tomography) and experimentally achieved classical correlations, quantum mutual information and discord.

\subsubsection{Fidelity of experimental density matrices}

\begin{figure}[h]
    	\centering
    	\begin{subfigure}{0.5\textwidth}
    	\includegraphics[width = 0.9\textwidth]{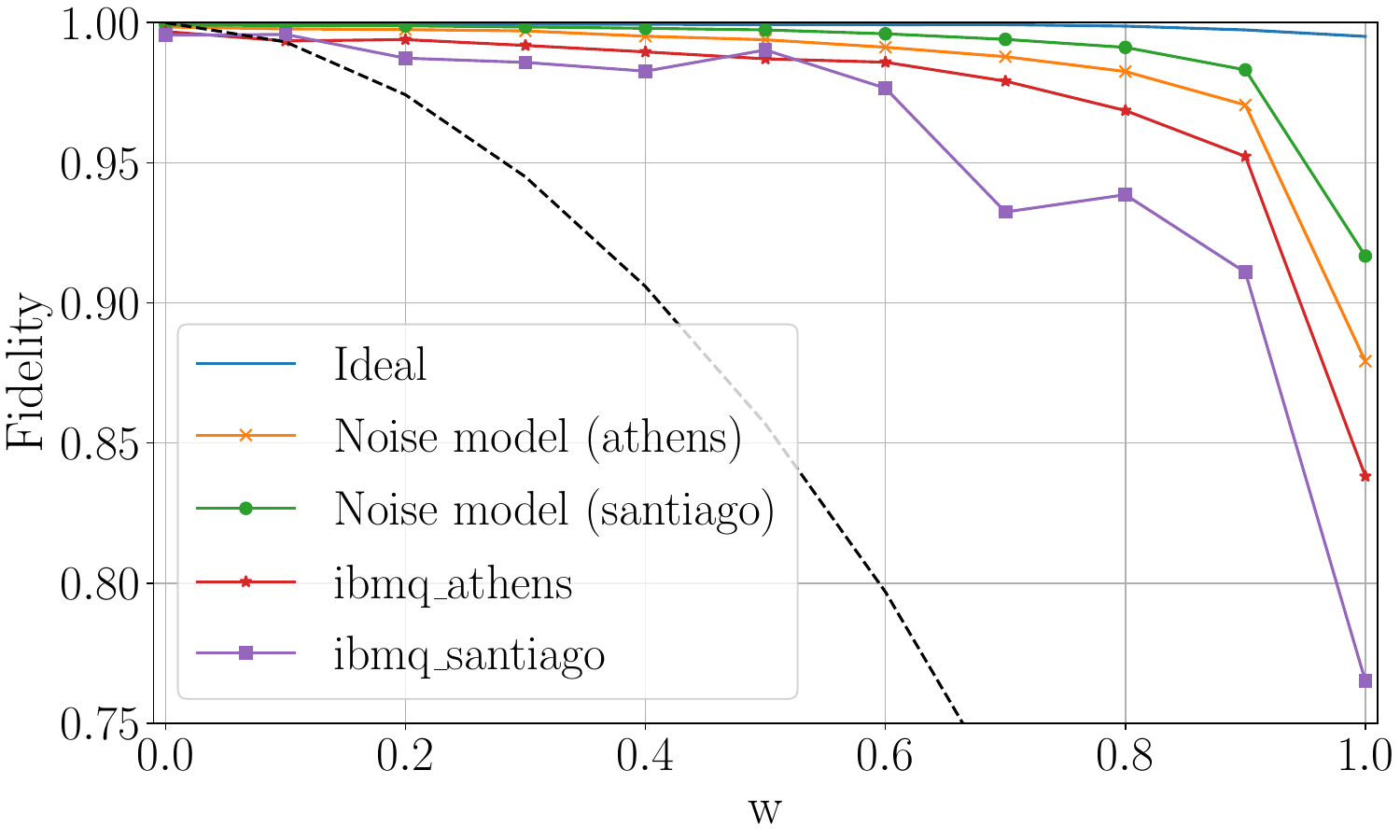}
    	\end{subfigure}
    	\begin{subfigure}{0.5\textwidth}\includegraphics[width = 0.9\textwidth]{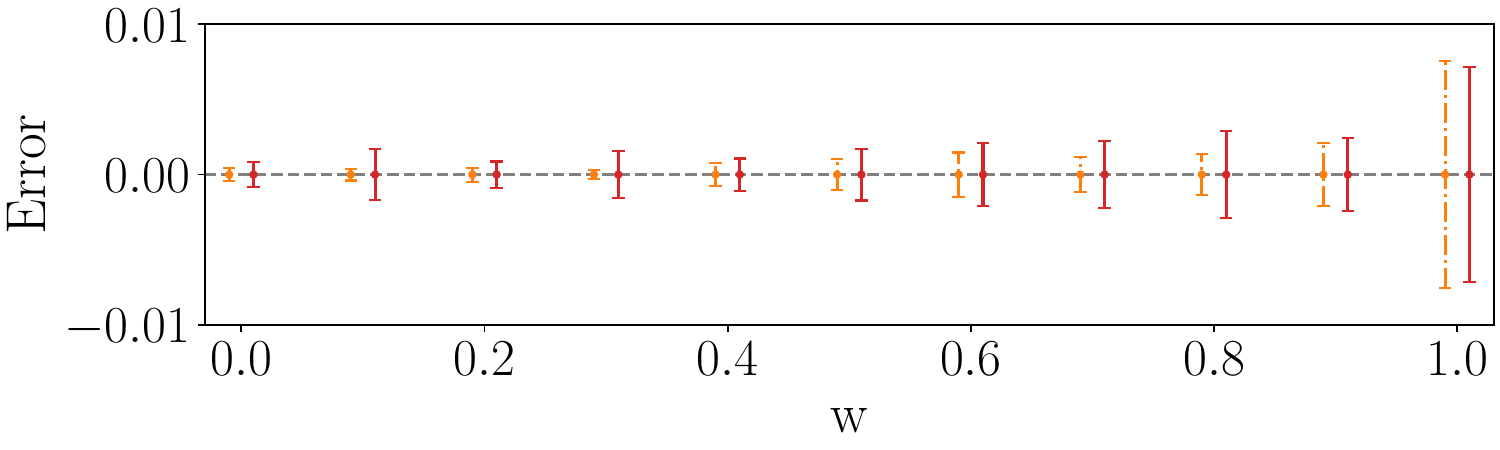}
    	\end{subfigure}
    	\caption{{\small (Above) Fidelity \eqref{eq:fid} of experimental Werner state density matrices reconstructed by Qiskit on \texttt{ibmq\_athens} and \texttt{ibmq\_santiago} for 5000 shots, and compared with the results from Qiskit simulator and noise models provided by Qiskit. The black dashed line corresponds to the extreme worst case where an identity matrix would be produced by the simulations (cf. eqs. (\ref{eq:ibmq_fid}) and its discussion). (Below) Standard deviation of fidelity over 10 simulations for \texttt{ibmq\_athens} (red solid line) and its noise model (yellow dash-dotted line)}}    	
    	\label{fig:fidelity}
\end{figure}

\Cref{fig:fidelity} shows the fidelity in the experiment using 
the quantum circuit on Fig. \ref{fig:G-compact} on \texttt{ibmq\_athens} and \texttt{ibmq\_santiago} with 5000 shots. This is compared to the ideal noiseless simulation 
with \textit{qasm-simulator}, and also the one using a
Qiskit noise model based on the properties of each real hardware.
The density matrix reconstructed using both \texttt{ibmq\_athens} and \texttt{ibmq\_santiago} with 5000 shots is close to the ideal one for small $w$, proving the performance of the real quantum computer in the corresponding domain, although it drops below $85\%$ and $75\%$, respectively, for $w=1$. At the time, this result suggests that it is necessary to improve the current noise model to describe the fidelity drop in a more faithful way. In this run, we see that slightly higher fidelity was obtained by \texttt{ibmq\_athens} compared to \texttt{ibmq\_santiago}.

\begin{figure}[h]
    	\centering
    	\includegraphics[width = 0.3\textwidth, trim = {2cm, 3cm, 2cm, 3cm}]{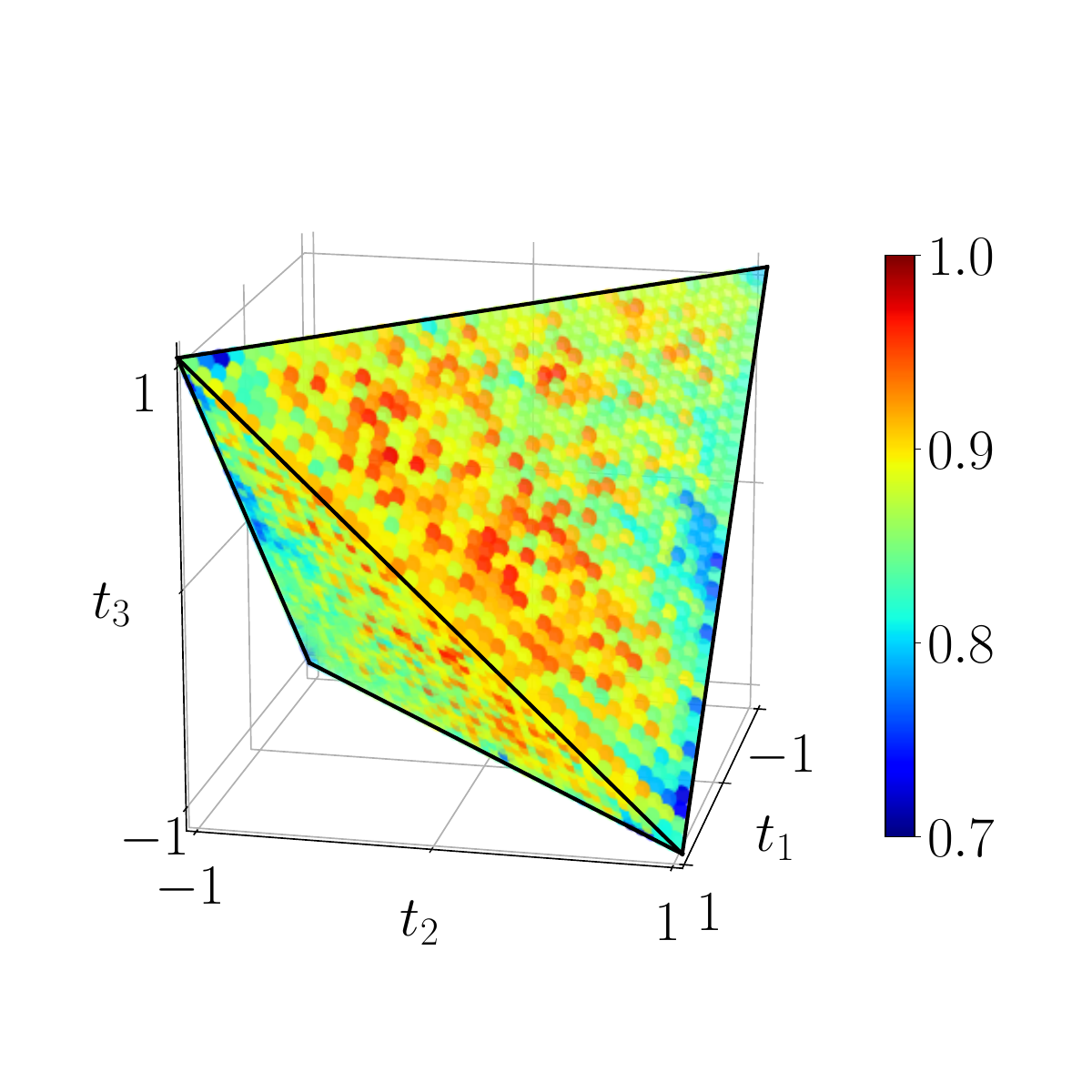}
    	\caption{{\small Experimental fidelity of BDS, scale $[0.7,1]$.}}    	
    	\label{fig:fidelity-tetrahedron}
\end{figure}

\Cref{fig:fidelity-tetrahedron} shows the fidelity of states on the full BDS tetrahedron, as computed from
the density matrices reconstructed from experiments on the \texttt{ibmqx2} backend, running the
circuit of \cref{fig:G-compact} with 1000 shots per measurement. We see that the fidelity is fairly high, $\sim 0.9$, over the whole domain which allows us to proceed with the calculations of quantum correlations in the following.

\subsubsection{Experimental classical correlations, quantum mutual information and discord}

Several quantities are studied on the real quantum computer. 
Again, we use the \texttt{ibmqx2} backend, running the circuit of \cref{fig:G-compact} with 1000 shots per measurement.

\begin{figure}[h]
		\centering
		\includegraphics[width=0.3\textwidth, trim = {2cm, 3cm, 2cm, 3cm}]{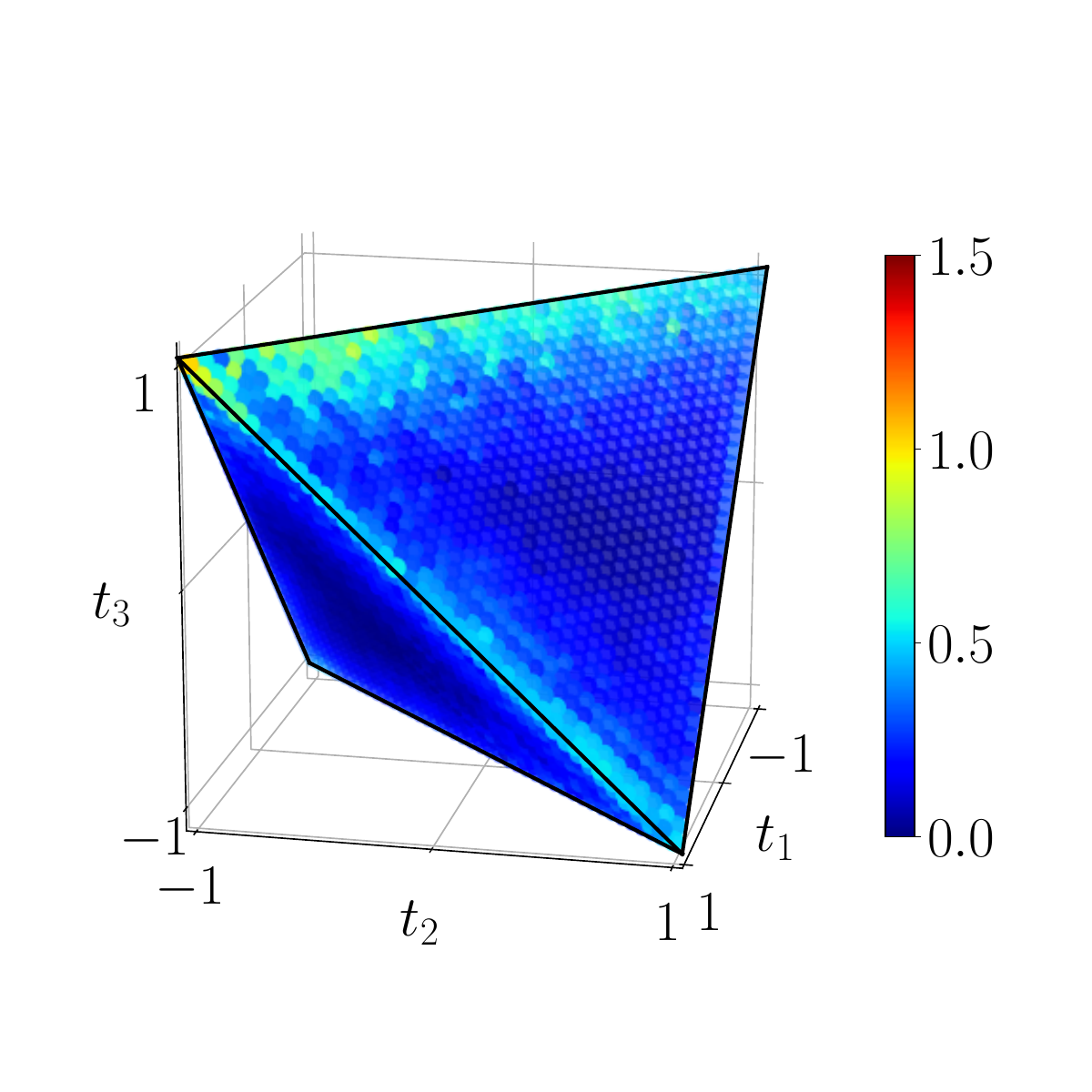}
		\caption{{\small Experimental classical correlations $\mathcal{C}_{BDS}$, scale $[0,1.5]$.}}
		\label{fig:Exp_cc_2}
\end{figure}

\begin{figure}[h]
		\centering
		\includegraphics[width=0.3\textwidth, trim = {2cm, 3cm, 2cm, 3cm}]{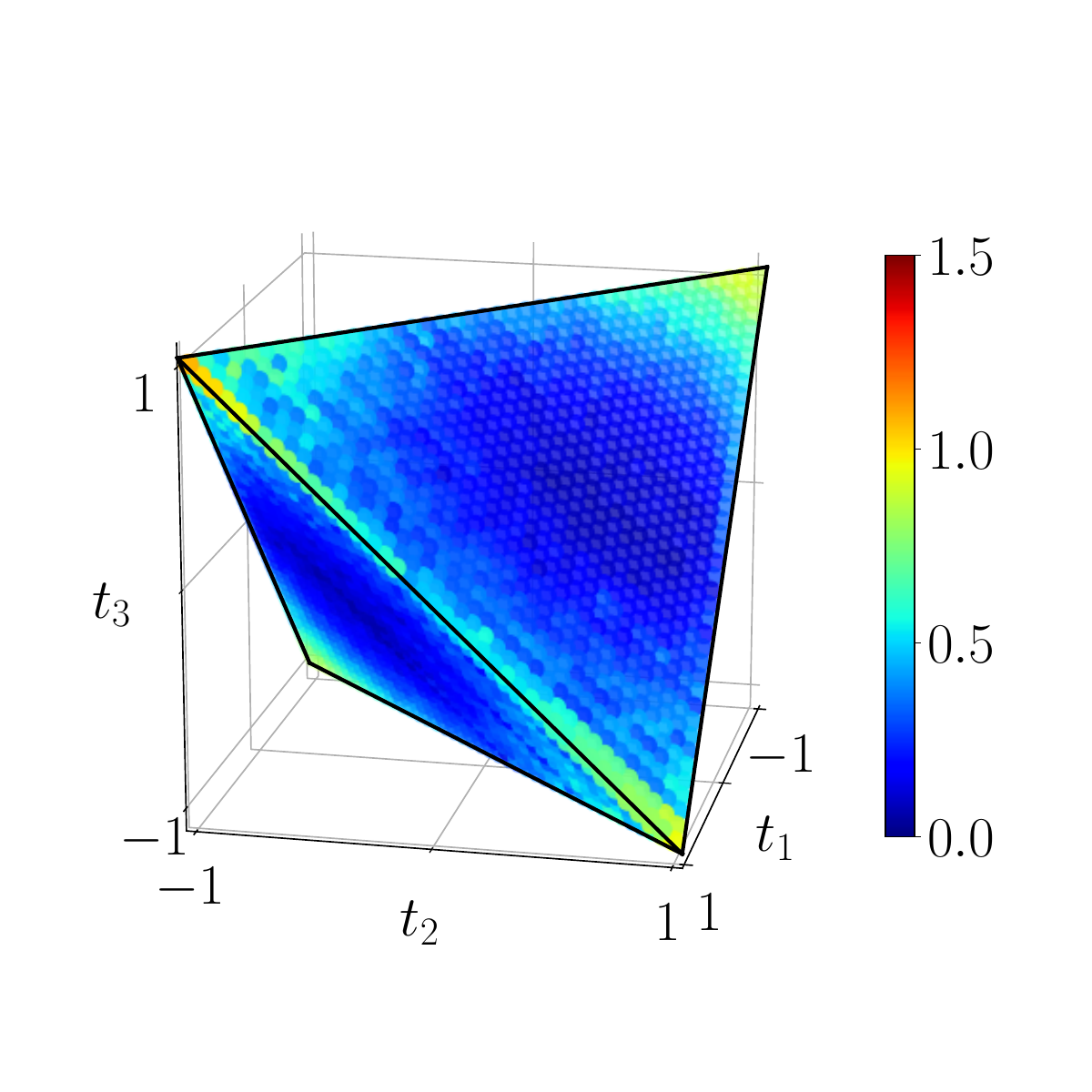}
		\caption{{\small Experimental quantum mutual information $\mathcal{I}_{BDS}$, scale $[0,1.5]$.}}
		\label{fig:Exp_qmi_2}
\end{figure}

\begin{figure}[h]
		\centering
		\includegraphics[width=0.3\textwidth, trim = {2cm, 3cm, 2cm, 3cm}]{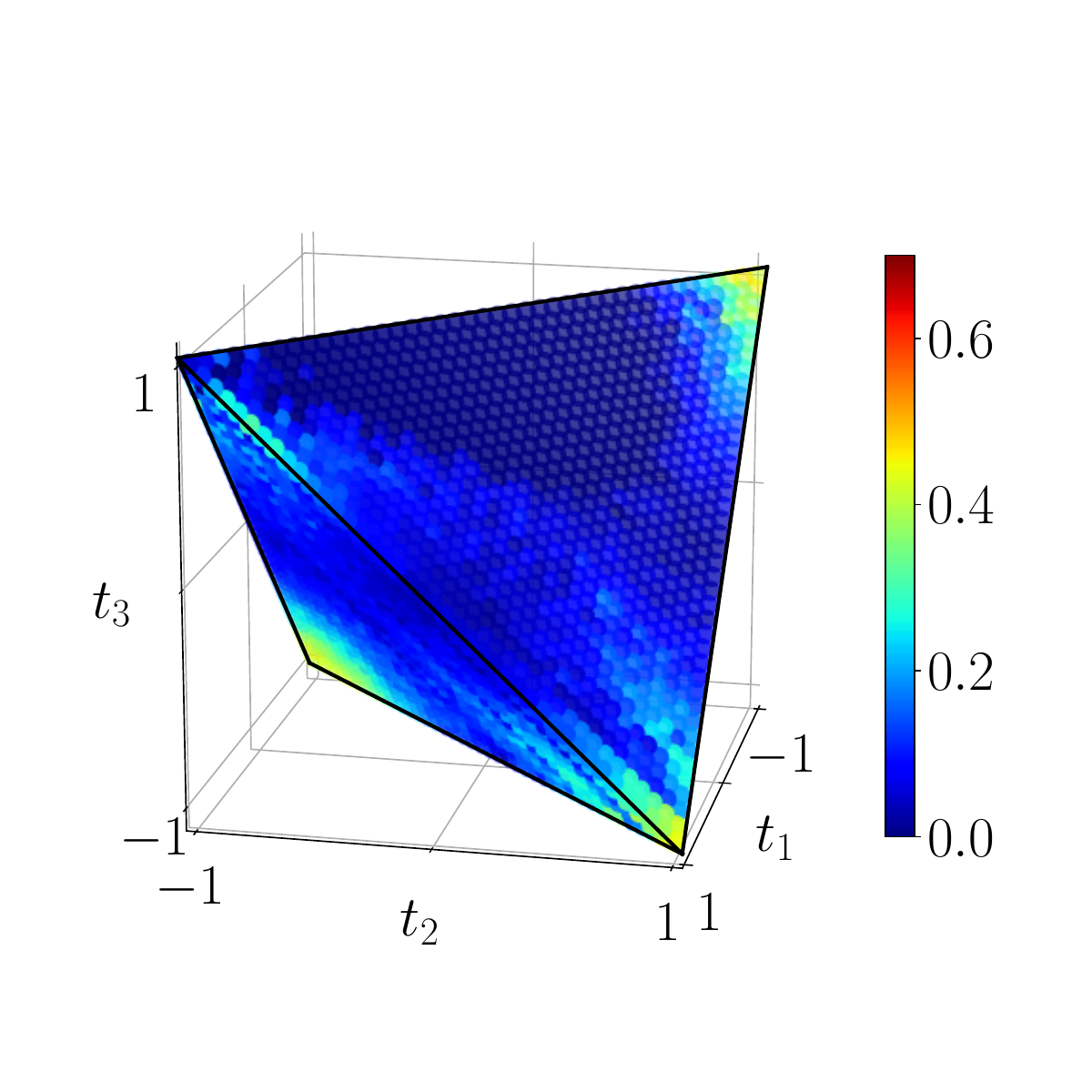}
		\caption{{\small Experimental discord $\mathcal{D}_{BDS}$ on the scale $[0,0.7]$.
		The three-pointed star pattern is barely visible.}}
		\label{fig:Exp_discord_2}
\end{figure}

One can see on \cref{fig:Exp_cc_2} that the experimental classical correlations seem to follow the theoretical predictions, nevertheless exhibiting lower values, especially visible on the edges of the tetrahedron. Quantum mutual information (\cref{fig:Exp_qmi_2}) seems to suffer the most of the lack of fidelity, indeed, its maximal values are nearly one unit below the theoretical ones. Finally, discord, plotted on \cref{fig:Exp_discord_2}, also decreased compared to the theory.

It can be noticed that classical correlations and quantum mutual information reveal an asymmetric behavior in the tetrahedron. The Bell states $\ket{\beta_{00}}$ and $\ket{\beta_{01}}$ both have higher values of these quantities than $\ket{\beta_{10}}$ and $\ket{\beta_{11}}$. This asymmetry is however also apparent in the noise simulations (see figs. \ref{fig:noisy-qmi} and \ref{fig:noisy-discord}) and thus seems to be explained by the noise model. More work would be needed to track the possible source of this asymmetry at the circuit level (with respect to the preparation of Bell states and edges of the tetrahedron).

	\section{Conclusions}
	
	In the preceding work, we have proposed new quantum circuits for the preparation of the entire class of Bell-diagonal states,
	and in particular Werner states, and tested them in simulations as well as on a real quantum device. To the best of our knowledge, they are the first correct, special-purpose BDS preparation circuits
	to be described. Furthermore, we have given a comprehensive reexamination of the central role of Bell-diagonal states in the study of entropic measures of quantum correlations, in particular quantum discord for which we found a specific equivalence with ``asymmetric relative entropy of discord''. More generally, and as a by-product of this work, we also found the remarkable general inequality \eqref{eq:ineqAsDiscords} between these two quantities: for any quantum state the former never exceeds the latter! We have illustrated the behavior of these measures on the BDS tetrahedron and the Werner line, comparing theory, circuit simulations and experiments.

	Currently, two primary qubits and two ancillary qubits seem necessary to prepare BDS on physical IBM Q devices; we recommend the circuit of \cref{fig:four-qubit} combined with \cref{fig:G-compact} for this purpose. However, this is not a fundamental restriction, but rather a consequence of the current limited capabilities of the hardware. On future quantum computers that support post-measurement gates and classically controlled measurement operations, the ancillary qubits will not be necessary. This highlights the value of developing not only the quantum systems themselves, but also the classical interfaces controlling them. In addition, fully-fledged classical control will make it possible to implement key protocols that rely on classical communication, such as quantum teleportation.
	
	We point out that it would be interesting (in future work) to implement the four qubit circuit template of \cref{fig:four-qubit-ancillary}. Indeed as explained in \cref{sect:circuits} it implements the unread measurements as an interaction with two environmental qubits. This could in practice
	be at an advantage compared to that of \cref{fig:four-qubit}, especially so in combination with \cref{fig:G-compact}. Indeed it places weaker demands on the topology of the underlying device. The degree to which this is true will of course depend on the particular device, but we speculate that \cref{fig:four-qubit-ancillary} will typically be at an advantage because it concentrates most operations to the qubits $a$ and $b$, and in particular, reuses the CNOT channel $a \to b$ where \cref{fig:four-qubit} requires an additional CNOT channel $c \to d$. 
	
	We have shown that the IBM Q devices allow for an experimental investigation of a large portion of the Hilbert space of two qubit systems, in particular for the correlation measures over the whole tetrahedron of BDS in \cref{sect:IBMQ_results}. The comparison of experiments and noisy simulations seems to show that the backend noise models provided by Qiskit are too optimistic, this reveals especially near corners of the tetrahedron.
	This is also visible at the level of the fidelity on the Werner line.

	\begin{acknowledgments}
		\noindent 
		The work of C. L. Chan was supported by grant no 200021 156672 of the Swiss National Science Foundation.
		We acknowledge use of the IBM Quantum Experience for this work.  The views expressed are those of the authors and do not reflect the official policy or position of IBM or the IBM Quantum Experience team.
		
	\end{acknowledgments}
	
	\appendix
	
	\section{Analytical solution for parameters}
	\label{app:analytical-solution}
	
	We provide here a full analytical solution to eqs.~\eqref{eq:compact-probs}.
	The procedure is implemented in the accompanying software \cite{gitlab-repo}.
	For brevity, we let $c_{\theta} = \cos(\frac{\theta}{2})$, $s_\theta = \sin(\frac{\theta}{2})$
	and $a_{jk} = \sqrt{p_{jk}}$. We then want to solve
	\begin{equation}
	\begin{array}{c}
	\begin{IEEEeqnarraybox}{lCl}
	a_{00} &=& c_{\alpha} c_{\beta} c_{\gamma}
	+ s_{\alpha} s_{\beta} s_{\gamma} \\
	a_{01} &=& c_{\alpha} c_{\beta} s_{\gamma}
	- s_{\alpha} s_{\beta} c_{\gamma} \\
	a_{10} &=& c_{\alpha} s_{\beta} c_{\gamma}
	- s_{\alpha} c_{\beta} s_{\gamma} \\
	a_{11} &=& c_{\alpha} s_{\beta} s_{\gamma}
	+ s_{\alpha} c_{\beta} c_{\gamma}
	\end{IEEEeqnarraybox}
	\end{array}
	\label{eq:compact-probs-abbr}
	\end{equation}
	for $\alpha$, $\beta$ and $\gamma$.
	First, inspecting the circuit of \cref{fig:G-compact}, we notice that the parameter $\alpha$
	controls the degree of entanglement of the output state. Guided by this, we isolate $\alpha$
	by computing the measure of entanglement
	\begin{IEEEeqnarray*}{rCl}
		\mqty|a_{00} & a_{01} \\ a_{10} & a_{11}| &=& a_{00} a_{11} - a_{01} a_{10} \\
		&=& (c_\alpha^2 + s_\alpha^2) (c_\beta s_\beta c_\gamma s_\gamma - c_\beta s_\beta c_\gamma s_\gamma) \\
		&& {} + c_\alpha s_\alpha (c_\beta^2 c_\gamma^2 + c_\beta^2 s_\gamma^2
		+ s_\beta^2 c_\gamma^2 + s_\beta^2 s_\gamma^2) \\
		&=& c_\alpha s_\alpha (c_\beta^2 + s_\beta^2) (c_\gamma^2 + s_\gamma^2) \\
		&=& c_\alpha s_\alpha = \frac{1}{2} \sin(\alpha). \yesnumber
	\end{IEEEeqnarray*}
	Choosing $\alpha \in \qty[-\frac{\pi}{2}, \frac{\pi}{2}]$, implying $\cos(\alpha) \geq 0$
	and also $c_\alpha \geq 0$, we thus get
	\begin{equation}
	\alpha = \arcsin\!\qty\big(2(a_{00} a_{11} - a_{01} a_{10})).
	\end{equation}
	Now, with knowledge of $c_\alpha$ and $s_\alpha$, eqs.~\eqref{eq:compact-probs-abbr}
	turn into a linear system of equations in four unknowns $c_\beta c_\gamma$, $c_\beta s_\gamma$,
	$s_\beta c_\gamma$ and $s_\beta s_\gamma$. The solution is
	\begin{equation}
	\begin{array}{c}
	\begin{IEEEeqnarraybox}{lCl}
	c_\beta c_\gamma &=& \frac{1}{\cos(\alpha)} ( c_\alpha a_{00} - s_\alpha a_{11}) \\
	c_\beta s_\gamma &=& \frac{1}{\cos(\alpha)} ( c_\alpha a_{01} + s_\alpha a_{10}) \\
	s_\beta c_\gamma &=& \frac{1}{\cos(\alpha)} ( s_\alpha a_{01} + c_\alpha a_{10}) \\
	s_\beta s_\gamma &=& \frac{1}{\cos(\alpha)} (-s_\alpha a_{00} + c_\alpha a_{11})
	\end{IEEEeqnarraybox}
	\end{array}
	\end{equation}
	where we have used $c_\alpha^2 - s_\alpha^2 = \cos(\alpha)$.
	
	In the case of $\cos(\alpha) = 0$, i.e.~$\alpha = \pm \frac{\pi}{2}$, we have
	$c_\alpha = \pm s_\alpha = \frac{1}{\sqrt{2}}$
	and the system \eqref{eq:compact-probs-abbr} is singular because the output state
	in fact depends only on $\beta \mp \gamma$. Then we may choose $\gamma = 0$ and find
	$\beta$ directly from \eqref{eq:compact-probs-abbr} through
	\begin{equation}
	c_\beta = \sqrt{2} a_{00}, \qquad s_\beta = \sqrt{2} a_{10}.
	\end{equation}
	(In fact, for $\cos(\alpha) = 0$, one of the parameters $a_{jk}$ must be
	negative unless $\beta = \gamma$, so only the case $\beta = \gamma = 0$
	is of interest to us.)
	
	If $\cos(\alpha) \neq 0$, we are able to construct the matrix
	\begin{equation}
	A \equiv \mqty(
	c_\beta c_\gamma & c_\beta s_\gamma \\
	s_\beta c_\gamma & s_\beta s_\gamma)
	= b c^\transp
	\end{equation}
	where we have defined the unit vectors
	\begin{equation}
	b = \mqty(c_\beta \\ s_\beta) \qq{and} c = \mqty(c_\gamma \\ s_\gamma).
	\end{equation}
	The matrices $A A^\transp = b c^\transp c b^\transp = b b^\transp$ and
	$A^\transp\! A = c b^\transp b c^\transp = c c^\transp$ are the projectors onto
	$b$ and $c$ respectively, and may be used to find $b$ and $c$ up to a sign (for
	example, $b = \pm \frac{A A^\transp x}{\norm{A A^\transp x}}$ for an arbitrary
	vector $x$ not orthogonal to $b$).
	We may fix the sign of $b$, e.g.~by imposing $c_\beta \geq 0$ (or $s_\beta \geq 0$ if
	$c_\beta = 0$); the sign of $c$ is then fixed by the condition $b c^\transp = A$.
	Finally, $\beta$ and $\gamma$ are determined by $b$ and $c$.

\section{Quantum implementation of classical operations}
\label{app:Q-implementation-classical-op}

We showed in \cref{sec:two-qubit} how we could reduce the number of quantum bits from four to two
by introducing classical registers and operations. However, current hardware does not support these.
Therefore, as well as for general theoretical interest, it is relevant to ask how the classical
parts of a circuit, including measurements, can be recast as purely quantum operations.

Any purely classical computation can be done on a quantum computer
using only the subset of quantum gates that map computational basis states to computational basis
states (that is, gates whose matrix representation have only one nonzero element in each row
and each column). For this to work, one must first arrange for the classical computation to
be reversible, which is always possible but may require ancillary bits
\cite[chap.~3.2.5]{nielsen_chuang_2010}.

It is also possible to rearrange a quantum circuit where decisions are taken based
on measurement outcomes into one where all measurements occur at the end of the
computation. One simply replaces each measurement with a CNOT operation targeting
a new qubit, and any gates conditional on the result being $1$ with the corresponding
controlled gates. This is the \emph{principle of deferred measurement}
\cite[chap.~4.4]{nielsen_chuang_2010}.

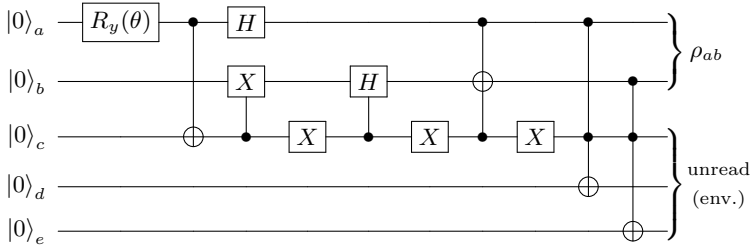
\begin{figure}[h]
	\centering
	\vspace{0.4cm}
	\[ \Qcircuit @C=1em @R=1em {
	\lstick{\ket{0}_a} & \gate{R_y(\theta)} & \ctrl{2} & \gate{H} & \qw & \qw & \qw
	& \ctrl{1} & \qw & \ctrl{2} & \qw & \qw \\
	\lstick{\ket{0}_b} & \qw & \qw & \gate{X} & \qw & \gate{H} & \qw & \targ
	& \qw & \qw & \ctrl{1} & \qw \\
	\lstick{\ket{0}_c} & \qw & \targ & \ctrl{-1} & \gate{X} & \ctrl{-1} & \gate{X}
	& \ctrl{-1} & \gate{X} & \ctrl{1} & \ctrl{2} & \qw \\
	\lstick{\ket{0}_d} & \qw & \qw & \qw & \qw & \qw & \qw & \qw & \qw & \targ & \qw & \qw \\
	\lstick{\ket{0}_e} & \qw & \qw & \qw & \qw & \qw & \qw & \qw & \qw & \qw & \targ & \qw
	\outputgroupv{1}{2}{12}{.6em}{1.3em}{\hspace{-.8em}\rho_{ab}}
	\outputgroupv{3}{5}{12}{.6em}{2.0em}{\hspace{.2em}
	\scriptsize\eqbox{c}{t}{unread\\(env.)}}
	} \] 	
	\caption{{\small Fully quantum version of the circuit of \cref{fig:werner-circuit}
	for preparing Werner states.}}
	\label{fig:werner-circuit-fully-quantum}
\end{figure}

\Cref{fig:werner-circuit-fully-quantum} shows the result of straightforwardly
applying the principle of deferred measurement to the circuit of
\cref{fig:werner-circuit}.
While the two circuits are theoretically equivalent, the rewriting has introduced
several extra quantum gates, including three Toffoli gates. The extra complexity
makes it likely that the purely quantum circuit is more sensitive to noise than
the one incorporating classical elements (though this statement of course depends on
the performance of the hypothetical in-circuit measurement, which we cannot in
fact assess). Therefore, we have not considered this circuit as a practical
alternative to the one in \cref{fig:werner-circuit}.
	
\section{Separability and entanglement of formation  for Werner states}\label{app:appendixwerner}

For the sake of completeness we summarize a few details for the determination of the separability threshold and  entanglement of formation. 

In the computational basis the Werner state reads
$$
\frac{1}{4}\begin{pmatrix}
1-w & 0 & 0 & 0\\
0 & 1+w & -2w & 0\\
0 & -2w & 1+w & 0\\
0 & 0 & 0 & 1-w
\end{pmatrix}
$$
and its partial transpose is given by
\begin{align*}
\rho^{T_B}_W = \frac{1}{4}
\begin{pmatrix}
1-w & 0 & 0 & -2w\\
0 & 1+w & 0 & 0\\
0 & 0 & 1+w & 0\\
-2w & 0 & 0 & 1-w
\end{pmatrix}
\end{align*}
The eigenvalues are easily calculated from the two $2\times 2$ blocks, and one finds
three degenerate eigenvalues equal to $\frac{1+w}{4}$ and another equal to $\frac{1-3w}{4}$. By the PPT criterion the state is separable if and only if $w\in [0, \frac{1}{3}]$ and entangled for $w\in (\frac{1}{3}, 1]$. 

To compute the entanglement of formation we apply the formulas of \cref{sec:entanglementmeasures}. First, we compute the matrix
$\tilde\rho_W=\sigma_2\otimes \sigma_2\rho_W^*\sigma_2\otimes\sigma_2$,
\begin{align*}
\tilde\rho_W = \frac{1}{4}
\begin{pmatrix}
1-w & 0 & 0 & 0\\
0 & 1+w & -2w & 0\\
0 & -2w & 1+w & 0\\
0 & 0 & 0 & 1-w
\end{pmatrix}
\end{align*}
and note that it is equal to $\rho_W$. Thus the square roots of the eigenvalues of $\rho_W\tilde\rho_W = \rho_W^2$ are given by the eigenvalues of $\rho_W$.
These are, in descending order, $\mu_1=\frac{1}{4}(1+3w)$, $\mu_2 = \mu_3 =\mu_4 = \frac{1}{4}(1-w)$.
Applying Wootters' formula \eqref{eq:conc}
we find  $C(\rho_W) = \max(0, \frac{3w -1}{2}$ and the entanglement of formation immediately follows from eq.~\eqref{eq:wootters}.

\bibliography{main.bib}
 
\end{document}